\documentclass[12pt,preprint]{aastex}

\received{}
\accepted{}
\journalid{}{}
\articleid{}{}
\slugcomment{To appear in {\it the Astrophysical Journal}}

\shortauthors{KOBAYASHI AND NOMOTO}
\shorttitle{LIFETIME OF TYPE IA SUPERNOVAE AND METALLICITY EFFECT}

\def\gtsim {>\kern-1.2em\lower1.1ex\hbox{$\sim$}~}   
\def\ltsim {<\kern-1.2em\lower1.1ex\hbox{$\sim$}~}   

\begin{document}

\title{The Role of Type Ia Supernovae in Chemical Evolution I:
Lifetime of Type Ia Supernovae and Metallicity Effect}
\author{Chiaki KOBAYASHI$^1$ and Ken'ichi NOMOTO$^{2,3}$}
\affil{$^1$ Mt. Stromlo Observatory, Research School of Astronomy \& Astrophysics, The Australian National University; Cotter Rd., Weston ACT 2611, Australia; chiaki@mso.anu.edu.au}
\affil{$^2$ Institute for the Physics and Mathematics of the Universe, 
University of Tokyo, Kashiwa, Chiba 277-8568, Japan}
\affil{$^3$ Department of Astronomy, School of Science,
University of Tokyo, Bunkyo-ku, Tokyo 113-0033, Japan;
nomoto@astron.s.u-tokyo.ac.jp}

\begin{abstract}
We construct a new model of Type Ia Supernovae (SNe Ia), based on the single degenerate scenario, taking account of the metallicity dependences of the white dwarf (WD) wind and the mass-stripping effect on the binary companion star.
Our model naturally predicts that the SN Ia lifetime distribution spans a range of $0.1-20$ Gyr with the double peaks at $\sim 0.1$ and $1$ Gyr.
While the present SN Ia rate in elliptical galaxies can be reproduced with the old population of the red-giants+WD systems, the large SN Ia rate in radio galaxies could be explained with the young population of the main-sequence+WD systems.
Because of the metallicity effect, i.e., because of the lack of winds from WDs in the binary systems, the SN Ia rate in the systems with [Fe/H] $\ltsim -1$, e.g., high-z spiral galaxies, is supposed to be very small.
Our SN Ia model can give better reproduction of the [($\alpha$, Mn, Zn)/Fe]-[Fe/H] relations in the solar neighborhood than other models such as the double-degenerate scenario.
The metallicity effect is more strongly required in the presence of the young population of SNe Ia.
We also succeed in reproducing the galactic supernova rates with their dependence on the morphological type of galaxies, and the cosmic SN Ia rate history with a peak at $z \sim 1$. 
At $z \gtsim 1$, the predicted SN Ia rate decreases toward higher redshifts and SNe Ia will be observed only in the systems that have evolved with a short timescale of chemical enrichment. This suggests that the evolution effect in the supernova cosmology can be small.
\end{abstract}

\keywords{galaxies: abundances --- galaxies: evolution --- stars: supernovae}

\section{Introduction}

The star formation histories of galaxies are imprinted on stellar populations of the galaxies. Because of the age-metallicity degeneracy for stellar populations, it is difficult to determine ``age'' from their colors and spectra alone. However, the elemental abundance ratios can provide independent information on ``age'', because different heavy elements are produced from different supernovae with different timescales \citep[e.g.,][]{pagel1997,matteucci2001}.
To derive the information of ``ages'' from elemental abundance patterns as a cosmic clock,
the most important uncertainty is the lifetime of Type Ia Supernovae (SNe Ia),
i.e., the evolutionary time (delay-time) of the progenitor from the main-sequence through the explosion, which is mainly determined by the lifetime of the relatively low-mass ($\ltsim 8 M_\odot$) companion star of the exploding white dwarf (WD).

There exist two distinct types of supernova explosions \citep[e.g.,][]{arnett1996,alex1997}; One is 
Type II supernovae (SNe II), which are the core collapse-induced explosions
of massive stars ($\gtsim \, 8M_\odot$) 
with short lifetimes of $10^{6-7}$ yrs,
and produce more $\alpha$-elements (O, Mg, Si, S, Ca, and Ti) 
relative to Fe with respect to the solar ratios (i.e., [$\alpha$/Fe]
$>0$).
Type Ib and Ic Supernovae (SNe Ib and Ic) are also the core-collapse supernovae, but have lost the envelope material before the explosions.
Recently, it is found that bright core-collapse supernovae, hypernovae (HNe), produce an important amount of Fe \citep[e.g.,][]{nom06}.
In chemical evolution models, the contributions of SNe Ib, Ic, and HNe can be included in that of SNe II \citep{kob06}.
The other is SNe Ia, which are the thermonuclear explosions 
of accreting WDs in close binaries 
and produce mostly Fe and little $\alpha$-elements \citep[e.g.,][]{nom94,hil00}.
The progenitors of the majority of SNe Ia are most likely 
the Chandrasekhar (Ch) mass WDs (e.g., Nomoto, Iwamoto \& Kishimoto 1997 for a review; see also Nomoto et al. 2007),
although the sub-Ch mass models might correspond to some peculiar 
subluminous SNe Ia.
The early time spectra of the majority of SNe Ia are
in excellent agreement with the synthetic spectra of the Ch mass models,
while the spectra of the sub-Ch mass models are too blue to
be compatible with observations \citep{hof96,nug97}.

For the evolution of accreting C+O WDs toward the Ch mass,
two scenarios have been proposed; 
One is a double-degenerate (DD) scenario, i.e., merging of double C+O WDs 
with a combined mass surpassing the Ch mass limit
\citep{ibe84,web84}.
From the long-term observational search (SPY project) of the WD binaries for the DD systems, more than thousand WDs are found.
However, their combined masses are smaller than the Ch mass, except for a few candidates near the Ch mass \citep{nap07,gei07,tov07}.
Theoretically, it has been suggested that the merging of WDs leads to accretion-induced collapse rather than SNe Ia \citep{sai85,sai98}.
The other is a single-degenerate (SD) scenario,
i.e., the WD mass grows by accretion of hydrogen-rich matter via mass transfer from
a binary companion \citep[e.g., ][for reviews]{nom00a,liv01}.
Two progenitor systems have been found for red-giant (RG) and near main-sequence (MS) companions \citep[e.g., ][]{hac99u}.
The SD scenario has predicted the detection of hydrogen-rich circumstellar matter from the companion star for a certain class of SN Ia binary systems, and such circumstellar matter has been actually detected for some SNe Ia \citep[e.g., ][]{ham03,pat07}.

The lifetime of SNe Ia has been estimated from different aspects as follows, and these estimates conflict each other;
(1) {\it Population synthesis} ---
The lifetime distribution of SNe Ia has been predicted with the population synthesis models for the evolution of binary systems \citep[e.g.,][]{yun05,bel05,tut07}.
The predicted rate of SD systems in those models is much smaller than DD systems, and the majority of SNe Ia have the lifetime as short as $\sim 0.1$ Gyr \citep[e.g.,][]{rui95,yun98}.
However, the optically thick winds from the WD, which were not taken into account in the above models, have been shown to play an essential role in the evolution of the accreting WDs \citep{hac96,hac99,hac99u}.
Furthermore, the stripping effect may increase the parameter space for the SD systems \citep{hac08}.
If these effects are taken into account in the population synthesis models, the DD rate could be smaller, and the SD rate could be larger.

(2) {\it Chemical evolution} ---
The typical lifetime of SNe Ia could have been constrained from the chemical evolution of the Milky Way Galaxy, most notably in the [$\alpha$/Fe]-[Fe/H] relations.
Metal-poor stars with [Fe/H] $\ltsim -1$ have [$\alpha$/Fe] $\sim 0.5$ 
for Mg, Si, and Ca on the average \cite[e.g.,][]{mcw95,cay04},
while disk stars with [Fe/H] $\gtsim -1$ show a decrease in [$\alpha$/Fe] 
with increasing metallicity \cite[e.g.,][]{edv93,ben04}.
Such an evolutionary change in [$\alpha$/Fe] against [Fe/H] 
has been explained with the early heavy-element production by SNe II and
the delayed enrichment of Fe by SNe Ia \citep{mat86}.
Conversely, chemical evolution models can constrain the nature of progenitor systems of SNe Ia;
for example, Yoshii, Tsujimoto, \& Nomoto (1996) 
estimated the lifetime of SN Ia progenitors to be as long as $0.5-3$ Gyr.

(3) {\it Supernova rates} ---
The redshift evolution of the SN Ia rate can give a rough estimate of the typical lifetime of SNe Ia (i.e., delay-time) for the assumed star formation history.
The decrease of the SN Ia rate seen at $z \sim 1$ corresponds to the delay-time of $\sim 3$ Gyr \citep{str04}.
However, the observed cosmic star formation rates (SFRs) involve uncertainties from dust extinction and completeness. 
The environmental dependence on galaxy evolution must be important if the SNe Ia rate depends on metallicity.
SN Ia rates depending on the galaxy type can give stronger constraint on the SN Ia progenitor models \citep{cal06}.
\citet{del05} suggested that the short lifetime of $\sim 0.1$ Gyr is required to explain the high SN Ia rate in radio-loud elliptical galaxies, and \citet{sul06} suggested the similar lifetime distribution from the relation between the SN Ia rate and the specific SFR.
Therefore, the bimodal distribution of the lifetime of SNe Ia, so-called ``A+B'' model, has been proposed \citep{sca05,man06}.

In addition, Kobayashi et al. (1998, hereafter K98) have found the metallicity effect on the WD winds, consequently, on the occurrence of SNe Ia, based on the simulations of the progenitor evolution by Hachisu et al. (1999ab).
If the iron abundance of the progenitors is as low as [Fe/H] $\ltsim -1.1$,
then the wind is too weak for SNe Ia to occur.
Thus the SN Ia rate in the system with [Fe/H] $\ltsim -1.1$ is supposed to be ``very small''.
This prediction has not been observationally confirmed yet, and several SNe Ia in metal-poor environment have been reported \citep[e.g.,][]{pri08}.
Statistical studies of a large sample are required,
since there must be a spread in the metallicity distribution function of stars in a galaxy and also the [O/Fe] is uncertain (see \S \ref{sec:zeffect} for the details).

If the lifetime distribution function of SNe Ia depends on metallicity, the detailed calculation of chemical evolution is even more important to understand the observations of supernova rates and abundance ratios.
The redshift evolution of the cosmic SN Ia rate is affected by chemical enrichment histories in various types of galaxies.
The estimates of the star formation histories of galaxies from their abundance ratios are affected by the metallicity effect on the SN Ia rate.
In addition, Fe can be produced by the four sources \citep{kob06}: i) SNe Ia, ii) relatively large Fe production from low-mass SNe II with $M=13-15M_\odot$, iii) large Fe production from HNe, or iv) the SN I.5 explosion of some AGB stars.
Therefore, not only [$\alpha$/Fe] but also other elements should be considered to discuss the contribution of SNe Ia.

This may be the case for dwarf spheroidal (dSph) galaxies; the observed low [$\alpha$/Fe] at low [Fe/H] in dSphs is often suggested to be due to the significant contribution of SNe Ia at such low metallicity \cite[e.g.,][]{ven04}.
However, there is a large variety among galaxies, and a decreasing trend in [Mg/Fe] at [Fe/H] $< -1$ may be seen in Draco \citep{she01} and Sculper \citep{she03}, but not for other $\alpha$ elements such as Si and Ca.
For more metal-poor stars, some stars show high [$\alpha$/Fe] \citep{koc08,fre09}, but others show low [$\alpha$/Fe] \citep{aok09,fre09b}.
Since [Mn/Fe] in dSph galaxies is as low as in Milky Way halo stars, the abundance pattern can be better explained with low-mass SNe II than SNe Ia \citep[][hereafter K06]{tol03,kob03,kob06}.
This is also supported by the low [Ba/Fe] in \cite{aok09}.
Chemodynamical evolution of dSph galaxies are complicated because of the low SFR and large dark matter content, and can be largely affected only by a few supernovae.
At least, the low [$\alpha$/Fe] in low metallicity stars does not necessarily
mean the contribution of SNe Ia, and certainly
more detailed studies are required under the reasonable model of SN Ia progenitors.

Similarly, it is argued that damped Lyman $\alpha$ (DLA) systems show low [$\alpha$/Fe] at low [Fe/H] \citep[e.g., ][]{pet99}.
\citet{pro02}, however, concluded that [Si/Fe] $\sim 0.3$ at [Fe/H] $< -1.5$, and also similar values were presented in \citet{des07}.
The elemental abundance ratios in DLA systems are affected by the dust depletion.
The dust depletion has been often estimated using [Zn/Fe] ratio, but there is no assurance that [Zn/Fe] $\sim 0$ for low-metallicity. In fact, an increasing trend of [Zn/Fe] has been observed in the Milky Way Galaxy \citep{pri00,nis07}.
\citet{nis07} showed that [S/Zn], which is free from the dust effect, of halo stars is similar to that in DLA systems.
Such [S/Zn] and [Si/Fe] could be explained by hypernovae.
Since the observation is for neutral gas in galaxies, it can be easily affected by dilution of hydrogen in chemodynamical simulations \citep{kob07}.
Again, more detailed studies are required.

The modeling of SNe Ia in the chemical evolution was first made by Greggio \& Renzini (1983), where the SN Ia rate is calculated from the distribution of binaries based on Whelan \& Iben (1973), and then extended by Greggio (1996) including the efficiency of accretion of the envelope of the secondary to the WD.
This model has recently been updated in \citet{gre05} and \citet{mat06}.
An alternative model was proposed by K98, fully based on the SD scenario, where the metallicity effect is essential to explain the [O/Fe]-[Fe/H] relation in the solar neighborhood. 
Kobayashi, Tsujimoto, \& Nomoto (2000, hereafter K00) applied this model for different types of galaxies, and succeeded in reproducing the present SN II and Ia rates in different types of galaxies.
Matteucci \& Recchi (2001, hereafter MR01), however, argued that they could not reproduce K98 results and that the lack of SNe Ia at low metallicities produces results at variance with the observed [O/Fe]-[Fe/H] relation.
However, this is due to their misunderstanding of the parameterization of the K98 models, as described in detail in \S 5.

In this paper, we first show the lifetime distribution functions derived from different SN Ia models. We then introduce our new SN Ia model by taking into account the stripping effect \citep{hac08} and show the lifetime distribution function as a function of metallicity (\S 2).
We then evaluate the SN Ia models from the comparison with observations.
The evolution of the elemental abundance ratios in the solar neighborhood (\S 3) and the galactic and cosmic supernova rates (\S 4) can be better reproduced with our SN Ia model including the metallicity effect.
In \S 5, we discuss the two formulations in K98 and MR01 in detail, and reproduce both results using the alternative formulation.
In \S 6, we summarize our conclusions.
The detailed discussion on the elemental abundance ratios in other galaxies will be in the second paper.

\section{Lifetime Distribution}
\subsection{Previous Models of Type Ia Supernovae}

Lifetime (i.e., delay-time) of SNe Ia is not a single parameter.
The SN Ia lifetime, $t_{\rm Ia}$, can be determined from the lifetime of companion stars in C+O WD binary systems, because the time interval from the start of mass accretion to the explosion is as short as $\ltsim 10^7$ yr.
Since the companion's mass ranges are different among various SN Ia scenarios,
the typical lifetime is different for the different SN Ia models.
In chemical evolution models, two formulations have been proposed in K00 and MR01 (see \S 5 for details, Eqs.[\ref{eqkob}] and [\ref{eqmat}]).
For the DD scenario, the timescale of gravitational radiation until the two WDs start to merge is also important, and the lifetime distribution functions have been provided from the population synthesis of the binary evolution.

\begin{figure}
\center
\includegraphics[width=18cm]{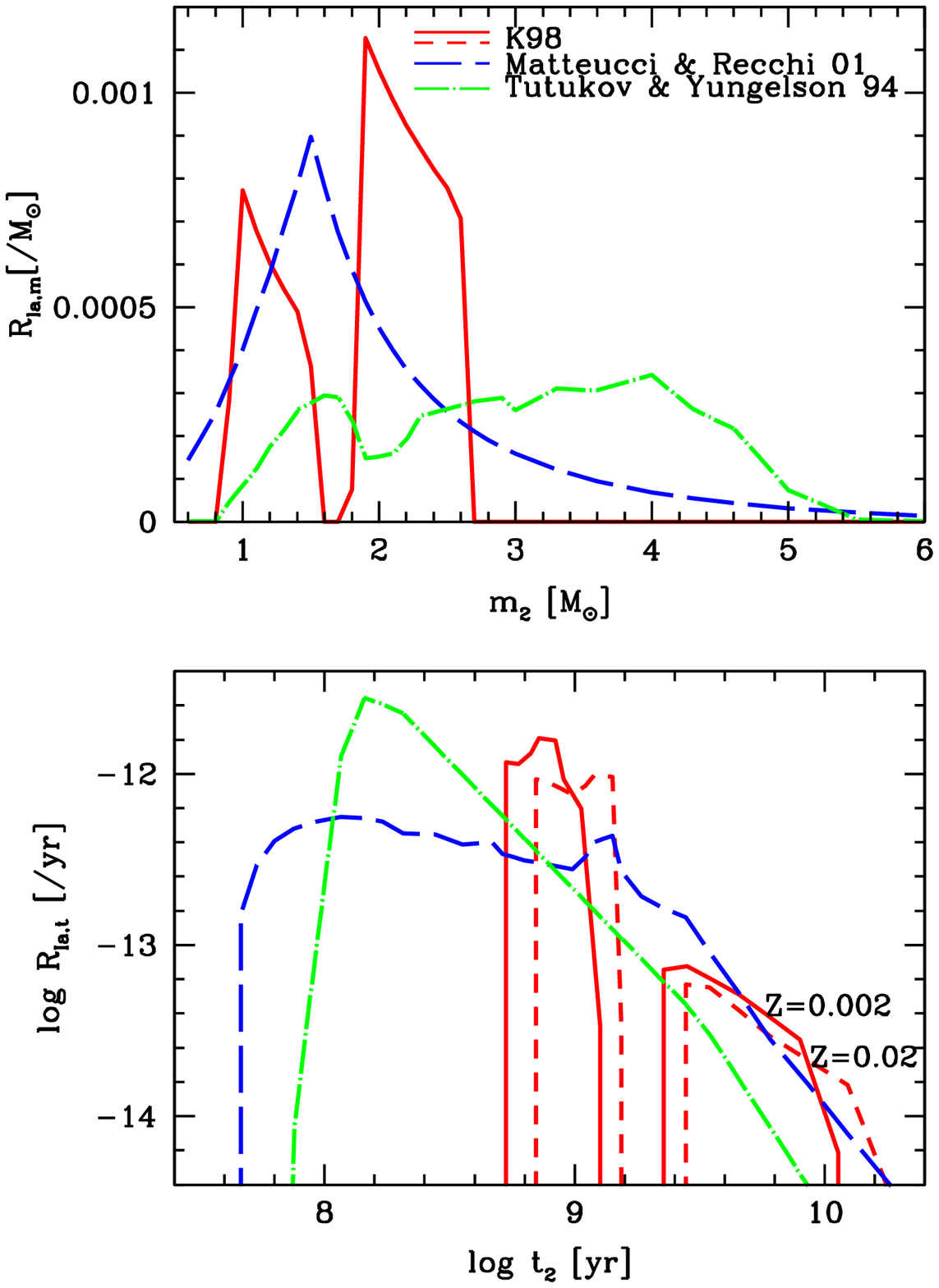}
\caption{\label{fig:life1}
The distribution functions of the progenitor mass (upper panel) and lifetime (lower panel), i.e., the SN Ia rates of stars with the same ages, for K98 model with $Z=0.002$ (red solid line) and $Z=0.02$ (red short-dashed line), MR01 model (blue long-dashed line, \citealt{mat01}), and DD model (green dot-dashed line, \citealt{tut94}).}
\end{figure}

Figure \ref{fig:life1} shows the distribution functions of the companion's mass $m_{\rm d}$ (upper panel) and lifetime $t_{\rm Ia}$ (lower panel) for the three SN Ia models.
These correspond to the SN Ia rate of the simple stellar population,
which is defined as a single generation of stars with the same age and metallicity.
We adopt the Salpeter initial mass function (IMF) $\phi(m) \propto m^{-1.35}$ for $m=0.07-50 M_\odot$, and the metallicity dependent main-sequence lifetimes, which is calculated by \citet{kod97} with the stellar evolution code described in \citet{iwan99}.

\begin{itemize}
\item The red solid and short-dashed lines are the distributions adopted in K98,which are calculated with Eq.[\ref{eqkob}] and the mass ranges at $Z=0.004$\footnote{The same as in Fig.1 of K00, but multiplied by the number fraction of WDs, $\int\frac{1}{m_1}\phi(m_1) dm_1=0.019$.}.
Because of the metallicity dependent stellar lifetimes, the resultant lifetime distribution $R_{{\rm Ia},t}$ slightly depends on metallicity
(solid lines, $Z=0.002$; short-dashed lines, $Z=0.02$).
The two components appear at $(m_{\rm d}, t_{\rm Ia}) = (\sim 2M_\odot,~0.5-1.5~{\rm Gyr})$ and $(\sim 1M_\odot,~2-20~{\rm Gyr})$, which correspond to the two types of binary systems, the MS+WD and RG+WD systems, respectively.
A kind of bi-modality, which is observationally proposed, has been naturally realized in our model.
Note that the spikes\footnote{In Fig.1 of K00, those are not seen due to a
coarse mass grid.} around $\sim 1$ Gyr are due to the change in
the stellar lifetime across the upper mass limit ($\sim 2
M_\odot$ depending of metallicity) for the occurrence of the He core
flash.

\item The green dot-dashed lines are for the DD scenario, where the \citet{tut94}'s function is adopted in our chemical evolution model.
The corresponding mass distribution (converted with the stellar lifetime function for $Z=0.02$) spans a range of $1-5M_\odot$\footnote{The drop of $R_{{\rm Ia},m}$ in the DD model is caused by the same reason as the spikes.}.
Because $\frac{d m_{\rm d}}{d t_{\rm Ia}}$ is larger for larger $m_{\rm d}$,
a typical lifetime is $t_{\rm Ia} \sim 0.1$ Gyr, the fraction with $\gtsim 1$ Gyr is very small, and the fraction with $\gtsim 10$ Gyr is almost zero.

\item The blue long-dashed lines are for the MR01-like model, calculated with Eq.[\ref{eqmat}], but with the same IMF as in our model.
The mass distribution (converted with the stellar lifetime function for $Z=0.02$) extends over $0-8 M_\odot$ having a peak at $\sim 1.5 M_\odot$.
The decrease of the rate toward larger $m_2$ is due to the IMF, and the decrease toward smaller $m_2$ is caused by $m_1+m_2 \ge 3 M_\odot$.
A typical lifetime is $t_{\rm Ia} \sim 0.3$ Gyr, which is longer than the DD model. 
The fraction with $\gtsim 1$ Gyr is two times larger than the DD model, and
the fraction with $\gtsim 10$ Gyr is small but not zero.
\end{itemize}
For each SN Ia progenitor model, the distribution function of the companion's mass and lifetime, i.e., the SN Ia rate in $1 M_\odot$ of simple stellar population is given in Tables \ref{tab:snia}-\ref{tab:snia3}.
In our SD scenario, the SN Ia rate is set to be 0 at [Fe/H] $< -1.1$.

\begin{deluxetable}{lccccc}
\tablenum{4}
\tablecaption{\label{tab:sniaparam}
Upper and lower mass of the binary companion of SN Ia progenitor systems depending on the iron abundance, in the mass unit of $M_\odot$.}
\footnotesize
\tablewidth{0pt}
\tablehead{
[Fe/H] & -1.1 & -1.0 & -0.7 & 0 & 0.4
}
\startdata
$m_{{\rm RG},\ell}$ & 0.9 & 0.9 & 0.9 & 0.9 & 0.8 \\
$m_{{\rm RG},u}$ &0.9 & 1.5 & 2.0 & 3.0 & 3.5 \\
$m_{{\rm MS},\ell}$ & 1.8 & 1.8 & 1.8 & 1.8 & 1.8 \\
$m_{{\rm MS},u}$ & 1.8 & 2.6 & 4.0 & 5.5 & 6.0 \\
\enddata
\end{deluxetable}

\subsection{A New SN Ia Model}

Hachisu et al. (2008, hereafter HKN08) introduced the stripping effect, where a part of the envelope mass of the companion stars is stripped by the interaction with the WD winds.
The efficiency is described by one parameter $c$, which depends mainly on the wind velocity. The effect of geometry can be given as a function of the mass ratio, and is included in $c$. The value of $c$ can be determined from the observed binary systems such as supersoft X-ray sources, and is $\sim 1.5-10$ for one source \citep{hac03} and will be $\sim 3$ on the average.

Although the binary evolution is calculated only for $Z=0.02$ in HKN08, the wind velocity depends on the metallicity, and the case with higher metallicity should be similar to the case with larger $c$ (Hachisu 2007, private communication). For $Z=0.002$, the mass stripping is supposed to be inefficient, and thus the mass ranges of companion stars are the same as in K98 and K00.
For higher metallicity, the mass ranges are wider.
Here we take the results with $c=0, 1,$ and $3$ for $Z=0.002, 0.004$, and $0.02$, respectively.
In the chemical evolution model, the upper and lower mass of the binary systems, $m_u$ and $m_\ell$, are given as a function of [Fe/H] (Table \ref{tab:sniaparam}), since iron is the most effective element for the opacity of the WD winds.
While the metallicity effect is introduced as the simple cut-off the SN Ia rate in K98 and K00, the metallicity effect is included in the parameter space of the binary systems that leads to SNe Ia in the present work.

\begin{figure}
\center
\includegraphics[width=18cm]{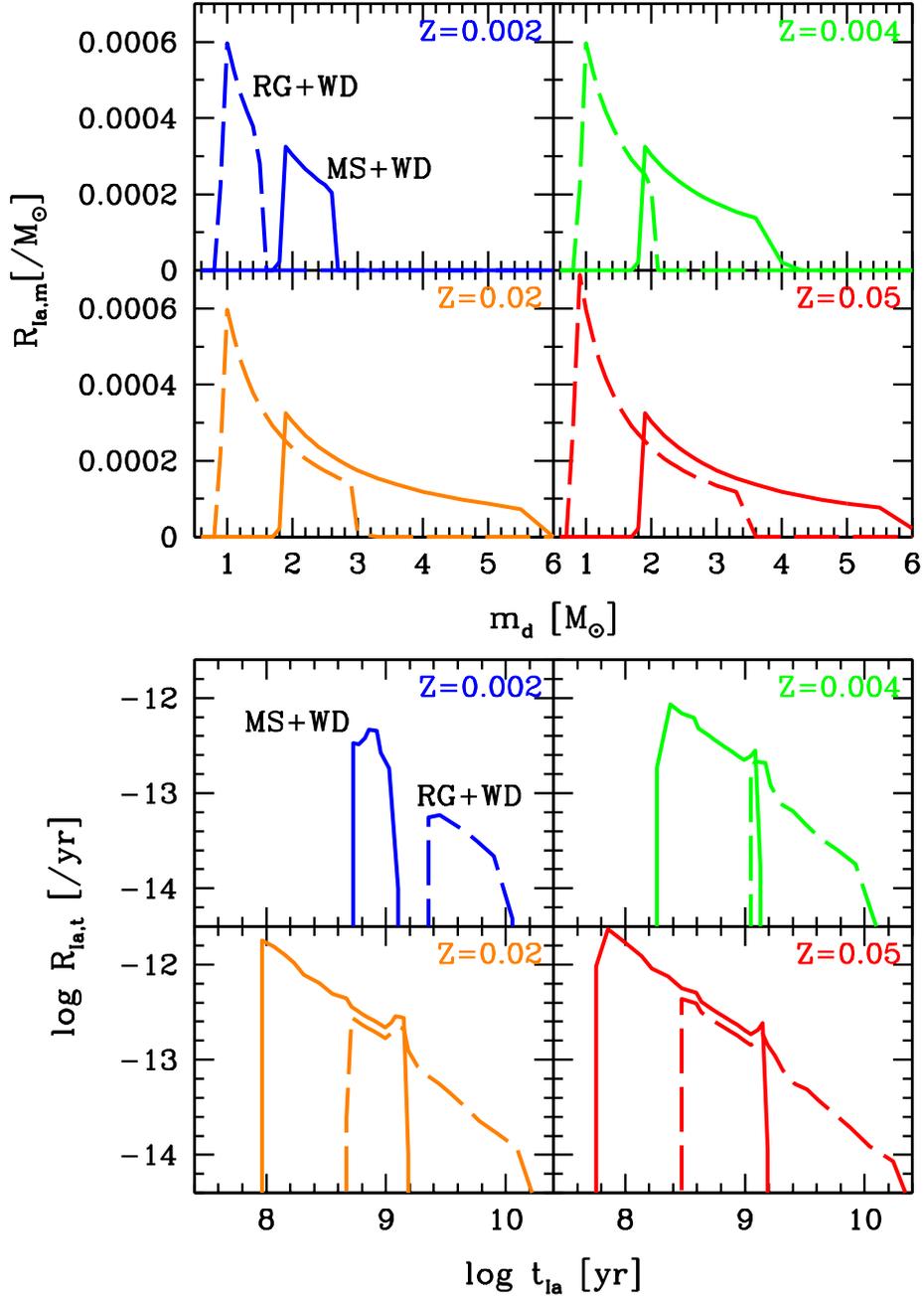}
\caption{\label{fig:life2}
The distribution functions of the progenitor mass (upper panel) and lifetime (lower panel) for the metallicity $Z=0.002$ (blue), $0.004$ (green), $0.02$ (orange), and $0.05$ (red).
The solid and dashed lines indicate the MS+WD and RG+WD systems, respectively, that correspond to the young and old population of SNe Ia.}
\end{figure}

Figure \ref{fig:life2} shows the SN Ia rate in the simple stellar population as functions of companion's mass $m_{\rm d}$ (upper panel) and lifetime $t_{\rm Ia}$ (lower panel) for the metallicity $Z=0.002$ (blue), $0.004$ (green), $0.02$ (orange), and $0.05$ (red).
The solid and dashed lines indicate the MS+WD and RG+WD systems that correspond to the young and old population of SNe Ia, respectively.
The total number of SNe Ia is determined to reproduce the chemical evolution in the solar neighborhood ([$b_{\rm RG}$, $b_{\rm MS}$]=[0.023, 0.023] at $Z=0.004$, see \S 3 for the details), and also adopted for the other types of galaxies (\S 4).
We should note that this function is different from HKN08 because the adopted IMF, stellar lifetimes, and binary parameters are different.
This function in Figure \ref{fig:life2} is better for chemical evolution models since it is calibrated to meet the observations in the solar neighborhood.

Figures \ref{fig:life3} show the summation of the young and old components of the SN Ia rate as functions of companion's mass (upper panel) and lifetime (lower panel).
The bimodal distribution is realized at 
$(m_{\rm d}, t_{\rm Ia}) = (\sim 2M_\odot,~0.1-0.5~{\rm Gyr})$ and $(\sim 1M_\odot,~1-3~{\rm Gyr})$ corresponding to the MS+WD and RG+WD systems, respectively.
The lifetimes at the two peaks depend on the metallicity: $t_{\rm Ia} \sim 0.5$ and $3$ Gyr for $Z=0.002$ (blue short-dashed line), $t_{\rm Ia} \sim 0.2$ and $1$ Gyr for $Z=0.004$ (green solid lines), $t_{\rm Ia} \sim 0.1$ and $1$ Gyr for $Z=0.02$ (orange dotted line) and $0.05$ (red long-dashed line).
Note that the spikes appears around $\sim 1$ Gyr because the stellar
lifetime is substantially longer for smaller mass stars that undergo
the He core flash.
This enhanced rate might cause the SN Ia driven galactic winds in 
elliptical galaxies.

At higher metallicity, the lifetime distribution extends to both shorter and longer lifetimes.
Such a short lifetime as $t_{\rm Ia} \sim 0.1$ Gyr is caused by the stripping effect that enables the accreting WDs with massive ($\sim 5-6 M_\odot$) companions to explode.
The lifetime as long as $t_{\rm Ia} \sim 20$ Gyr is caused by the elongation of the lifetime of small-mass companion stars, which is dominant in present-day elliptical galaxies.

HKN08 model predicts the existence of various types of
circumstellar matter (CSM) around the WDs just before making SN Ia
explosions depending on the mass accretion rates (and thus on the
binary parameters) at the last stages.
(1) case WIND: If the accretion rate is high enough for the optically
thick WD wind continue to blow, the stripped-off material forms CSM
very near the WD because the velocity of CSM is rather low; this case
corresponds to SNe 2002ic (e.g., Deng et al. 2003) and 2005gj
(Aldering et al. 2006), where SN Ia undergoes circumstellar
interaction.
(2) If the accretion rate has decreased below the critical rate for
the wind to occur at the explosion, the stripped-off material forms
CSM with various speed, but it has been dispersed far from the WDs.
Depending on the distance from the WDs and geometry of the materials,
CSM features may or may not be observed as in SN 2006X (Patat et
al. 2007) or SN 2007af (Simon et al. 2007). We also note that the
speed of the stripped-off materials is as low as $10-100$ km s$^{-1}$,
so that the sweeping effect of very fast wind discussed by Badenes et
al. (2007) is not expected.

\subsection{Metallicity Effect}
\label{sec:zeffect}

After our theoretical prediction of the metallicity effect, the SNe Ia in low-luminous galaxies have been paid attention.
K00 commented on SN1895B and SN1972E in NGC 5253, and SN1937C in IC 4182
that the metallicities of the host galaxies are higher than [Fe/H] $= -1.1$.
Hamuy et al. (2000) showed that no host galaxies are found at [Fe/H] $< -1$ (Fig. 1 in \citealt{ham01}) for 44 supernovae.
However,
SN1999aw is in the galaxy with $M_{\rm B}=-12.2$ \citep{str02}, which roughly corresponds to [O/H] $\sim -1.2$ \citep{tre04}.
SN2005cg is in the host galaxy with $M_{\rm r}=-16.75$ and [O/H] $=-0.4$ to $-0.1$ \citep{qui06}.
SN2007bk is in the galaxy with [O/H] $=-0.59$, but located at a distance of 8.7 kpc \citep{pri08}, where [O/H] is estimated as $\sim -0.7$ (see below for the details).
\citet{bad09} showed the stellar population near the supernova remnants (SNRs) of SNe Ia in the Small Magellanic Could from the color-magnitude diagram. The mean stellar metallicities are [Fe/H] $= -0.96, -0.94, -1.15$, and $-0.80$, but there is also a young metal-rich population as they commented.
From the statistical studies of a large sample such as the Sloan Digital Sky Survey and the Supernova Legacy Survey, no such metallicity effect has been seen in observations yet \citep[e.g.,][]{gal05,gal08,how09}, and rather opposite effect (i.e., decreasing of SN Ia rates toward higher metallicity) has been suggested by \citet{coo09}.

Note, however, that the following uncertainties are involved in these arguments;
i) The observation is the emission weighted gas phase [O/H], and can be different from the [Fe/H] of the progenitor system; gas metallicity can be larger then stellar metallicity in the chemically evolved and closed system, or can be smaller due to the dilution by gas accretion and/or inhomogeneous effects (i.e., stars tend to be formed from metal-enriched gas).
ii) The metallicity correlates with the galaxy mass (the mass-metallicity relation, e.g., \citealt{tre04}), but the scatter is larger for smaller galaxies.
iii) The mean stellar [$\alpha$/Fe] correlates with the galaxy mass in the sense that [$\alpha$/Fe] is larger for more massive galaxies ([$\alpha$/Fe] $\sim 0-0.3$, e.g., \citealt{tho03}), although the errorbar is quite large.
iv) There exits a radial gradient of metallicity, where the metallicity is lower in outside \citep[e.g.,][]{kob99}.
v) There exits a metallicity distribution of stars even at a certain location in galaxies (Fig. \ref{fig:mdf}).

\citet{foe08} reported that there is no radial dependence of SN Ia rate found in early-type galaxies.
However, this work also did not reach the metallicity threshold.
The metallicity gradient of early-type galaxies is typically $\Delta {\rm [Fe/H]}/\Delta\log r \sim -0.3$, and the metallicity at the effective radius is $<{\rm [Fe/H]}>_{\rm e} \sim -0.3$ to $0$ \citep{kob99}.
The metallicity effect would be seen at $r \gtsim 10 r_{\rm e}$ in intermediate-mass galaxies.
In dwarf galaxies, SNe Ia can be found at $r \sim 50$ kpc, because the effective radius does not become so small \citep{bin84}, and because the metallicity gradient is shallower for smaller galaxies \citep{spr09}.
For SN2007bk, the local metallicity is roughly estimated as [O/H] $\sim$ [Fe/H] $\sim -0.7$ with $\Delta{\rm [Fe/H]}/\Delta \log r/r_{\rm e} \sim -0.2$ and $r_{\rm e} \sim 1.5-2$ kpc.

\section{[X/Fe]-[Fe/H] Relations}
\label{sec:chem}

Using the one-zone chemical evolution model, we show the evolution of elemental abundance ratios in the solar neighborhood for different SN Ia models.
The star formation history is determined from the following observations in the solar neighborhood.
The formulation is described in K00; we use the model that allows the infall of primordial gas from outside the disk region.
The SFR is assumed to be proportional to the gas fraction.
For the infall rate, we adopt the same formula as in K98 and K06
that is proportional to $t\exp[-\frac{t}{\tau_{\rm i}}]$.
The input parameters are the Galactic age of $13$ Gyr and infall timescale $\tau_{\rm i}=5$ Gyr. The star formation timescale $\tau_{\rm s}=2.2$ Gyr is determined from the present gas fraction $f_{\rm g}=0.15$ (see K00 for the definition).
The same parameters are set for the other SN Ia models, because the contribution of SNe Ia on the gas fraction is negligible.
The nucleosynthesis yields as functions of mass, metallicity, and explosion energy are given in K06 for HNe, SNe II, and SNe Ia.
Photometric evolution is also calculated as in K00.

Figure \ref{fig:mdf} shows (a) the SFR, (b) the age-metallicity relations, and (c) the metallicity distribution function (MDF) for different SN Ia models.
The resultant SFR peaks at $t \sim 8$ Gyr, but does not very much change for $t \sim 5-13$ Gyrs, and the present SFR is consistent with the observational estimate.
The red solid, cyan dotted, green short-dashed, blue long-dashed, and magenta dot-dashed lines are for our new model with and without the metallicity effect, the DD, MR01-like, and K98 models, respectively.
All models can give excellent agreement with these observations.
Compared with the K98 model (magenta dot-dashed line), our new SN Ia model (red solid line) gives slightly slower increase in [Fe/H] after [Fe/H] $\sim -1$ (panel b), and slightly larger number of metal-poor stars around [Fe/H] $\sim -1$ (panel c).
This is because the metallicity effect is included not with a sharp cut-off but with a gradual decrease in our new model.
Recently, \cite{nord04} showed a narrower MDF than the plotted observations \citep{edv93,wys95}, which may suggest that a sharp cut-off exist in the SN Ia rate at [Fe/H] $\sim -1$.

The total number of SNe Ia is chosen to meet the MDF;
For the DD model, the total fraction of SNe Ia is set to be $0.001$.
For the MR01-like model, the binary fraction (see \S 5 for the definition) of $A=0.035$ is adopted.
In our models, the binary parameter $b$ denotes the fraction of primary stars that eventually explode as SNe Ia for the RG+WD and MS+WD systems. The combination of $b_{\rm RG}$ and $b_{\rm MS}$ can be determined from the metal-rich end of the MDF and the [O/Fe]-[Fe/H] relation at [Fe/H] $\gtsim -1$.
For our new SN Ia model, $b$ is normalized at $Z=0.004$, and [$b_{\rm RG}$, $b_{\rm MS}$]=[0.023, 0.023] are chosen for the best model. (Note that the models with [0.015, 0.028], [0.010, 0.032] and [0.005, 0.035] are also possible.)
Since the mass range expands for higher metallicity, the fraction of SNe Ia is 0.040 and 0.032 at $Z=0.05$ for the RG+WD and MS+WD systems, respectively.
For the case without the metallicity effect, the companion's mass ranges for $Z=0.02$ and [$b_{\rm RG}$, $b_{\rm MS}$]=[0.027, 0.027] are adopted.
For the K98 model, [$b_{\rm RG}$, $b_{\rm MS}$]=[0.02, 0.04] are chosen in K06 for the updated nucleosynthesis yields.

In the solar neighborhood, elemental abundance ratios are observed for a number of stars over a wide range of metallicity, which gives one of the most stringent constraints on the progenitors of SNe Ia.
Figures \ref{fig:xfe}-\ref{fig:xfe-c} show the evolutions of elemental abundance ratios [X/Fe] for O, Mg, Mn, and Zn, against the iron abundance [Fe/H].

In the early stage of the galaxy formation, only SNe II explode, and [$\alpha$/Fe] stays constant. Because of the delayed Fe production by SNe Ia, [$\alpha$/Fe] decreases toward 0.
The observations with UV OH lines show a monotonic increase in [O/H] towards the lower metallicity \citep{isr98,boe99}, but these are largely affected by the 3D effect of the atmosphere model and are not plotted here.
The O observations with the forbidden line and infrared lines are more reliable, and show the plateau rather than the increase as well as the other $\alpha$-elements (Mg, Si, Ca, and Ti).

Manganese shows a characteristic behavior.
Observational data show that [Mn/Fe] increases toward higher metallicity.
\citet{fel07} concluded that the Mn trend is likely to be due to the metallicity dependent SNe II, but the trend can be naturally reproduced by the delayed enrichment by SNe Ia since SNe Ia produce [Mn/Fe] $>0$ (K06).
This is consistent with that this increasing trend of [Mn/Fe] appears from the same [Fe/H] as the decreasing trends of [$\alpha$/Fe].
In the observation, a larger increase of [Mn/Fe] may be seen at [Fe/H] $\sim 0$ than our models, which may be due to the metallicity dependence of SN Ia yields, which is not included in our models.

The species of zinc depend on the metallicity.
At low metallicity, $^{64}$Zn is produced by complete Si-burning in HNe \citep{ume02}, which are assumed to be a half of massive ($\ge 20 M_\odot$) SNe II with the lifetime of $\ltsim 10^7$ yr (K06).
At high metallicity, neutron-rich isotopes of zinc $^{66-70}$Zn are produced by neutron-capture in He and C burning, which is larger for higher metallicity (Fig. 5 in K06) and ejected by SNe II.
Because of the combination of the lifetime and metallicity effects, [Zn/Fe] shows an interesting track.

\begin{itemize}
\item In the DD model (green short-dashed line),
[$\alpha$/Fe] decreases and [Mn/Fe] increases too early and too quickly from [Fe/H] $\sim -2$ compared with the observations.
This is because the lifetime of the majority of SNe Ia is shorter than $0.3$ Gyr (Fig. \ref{fig:life1}). 
The SN Ia lifetime is so short that [Zn/Fe] decreases by $0.25$ dex from [Fe/H] $\sim -2$ to $\sim -0.6$.
Because the SN Ia rate sharply decreases for longer lifetime, and because the SFR does not change very much, the contribution of HNe becomes larger than SNe Ia at later time. This results in the [Zn/Fe] increase from [Fe/H] $\sim -0.6$ to $\sim 0$, which is not seen in the observations.

\item In the MR01-like model reproduced with our code (blue long-dashed line),
the decrease in [$\alpha$/Fe] and the increase in [Mn/Fe] are seen at [Fe/H] $\sim -2$ as early as in the DD model.
The slope of [$\alpha$/Fe] and [Mn/Fe] against [Fe/H] is a little shallower than the DD model because the fraction of lifetime with $\gtsim 1$ Gyr is larger.
The [Zn/Fe] decreases only by $0.15$ dex from [Fe/H] $\sim -2$ to $\sim -0.6$.

\item Our SN Ia model with the metallicity effect (red solid line) gives the best agreement with the observations.
The decrease in [$\alpha$/Fe] and the increase in [Mn/Fe] are determined not from the lifetime effect but from the metallicity effect on the SN Ia rate.
Since the lowest metallicity to produce SNe Ia is [Fe/H] $=-1.1$, 
the evolutionary change is caused at [Fe/H] $\sim -1$ by
the companion stars with $M_{{\rm MS},u} = 2.6 M_\odot$ after the main-sequence lifetime of $\sim 0.5$ Gyr.
In the K98 model, the metallicity effect is included more sharply, and thus the evolutionary change appears a bit more sharply.

Our model predicts an interesting evolution of [Zn/Fe] around [Fe/H] $\sim -1$, as a contrast to other models;
[Zn/Fe] increases from [Fe/H] $\sim -1.5$ by the metallicity effect on Zn yield, and decreases from [Fe/H] $\sim -1$ by the contribution of SNe Ia.
This trend seems to be consistent with the observations, although the scatter is not small (\citet{nis04} at [Fe/H] $<-1$ and \citet{nis07} with the NLTE correction at [Fe/H] $>-1$).
Further observations of these elements at [Fe/H] $\gtsim -1.5$ is critically important in order to identify the SN Ia progenitors and to clarify the metallicity dependence on the nucleosynthesis yields.

\item If we do not include the metallicity effect (dotted line in Fig. \ref{fig:xfe2}), [$\alpha$/Fe] starts to decrease at [Fe/H] $\sim -2$.
This corresponds to the shortest SN Ia lifetime of $\sim 0.1$ Gyr for $M_{{\rm MS},u}=5.5 M_\odot$. This is too early to be compatible with the observations.
In the K98 model, the [$\alpha$/Fe] decrease starts at [Fe/H] $\sim -1.5$ (see Fig. 3 in K98), because the shortest lifetime ($\sim 0.5$ Gyr for $M_{{\rm MS},u}=2.6 M_\odot$) is longer than our new model.
In other words, without the metallicity effect, the existence of the young population of SNe Ia cause the [$\alpha$/Fe] decrease at much smaller [Fe/H], and the discrepancy with the observations becomes larger.

\item If there is the metallicity effect, these results do not depend very much on the $c$ parameter (Fig. \ref{fig:xfe-c}).
Even with $c=10$ that leads to the shortest lifetime of 0.03 Gyr, the difference is not visible.
This is because the majority of stars have low metallicity as $Z<Z_\odot$, and the metal-rich component can contribute only after $\sim 10$ Gyr.
If the stripping is effective even for $Z=0.002$ (long dashed line), the results does not change very much.
Therefore, the evolutionary change in the [X/Fe]-[Fe/H] relations is simply caused by the limit of the WD winds, as introduced in K98.

\end{itemize}

Figure \ref{fig:snsnr} shows the SN Ia rate history in the solar neighborhood.
Even with the same present SN Ia rate, different SN Ia models produce different SN Ia rate histories.
For the DD model, the lifetimes of majority of SNe Ia lifetime are short, and the SN Ia rate shows a rapid increase at $t<1$ Gyr and then a constant until the present time.
For the SD model, the lifetimes are a bit longer, and the SN Ia rate shows a slower increase at $t<5$ Gyr.
With the metallicity effect, SNe Ia start to occur at 3 Gyr when [Fe/H] reaches -1.1 (Fig.\ref{fig:mdf}b).
Compared with our previous work, our new model shows a more gradual increase due to the smoother SN Ia lifetime distribution as a function of metallicity.

In Figure \ref{fig:zeffect}, the supernova rate is plotted against the metallicity indicators: (a) the iron abundance, (b) the gas-phase metallicity, and (c) the mean stellar metallicity (see Eq.[15] in K00 for the definition).
Since the SN Ia rate is assumed to be 0 at [Fe/H] $< -1.1$, the rapid decrease is seen at [Fe/H] $\sim -1$ after the short time-delay of the chemical enrichment (panel a).
Because of the oxygen enhancement at lower metallicity, the decrease is seen at higher [O/H] as $\sim -0.6$.
In dwarf galaxies, however, the decrease will be seen at [O/H] $\sim -1$ since the observed dwarf galaxies have low [O/Fe].
For the mean metallicity (panel c), which is observed with the absorption lines of integrated stellar populations, the decrease is seen at lower $<{\rm [Fe/H]}>$ as $\sim -1.2$.
With the mass-metallicity relation of galaxies \citep{tre04},
the metallicity threshold corresponds to $M_{\rm B} = -15$ mag and $3 \times 10^8 M_\odot$.

\citet{man06} suggested that $\sim 50\%$ of SNe Ia should be composed of the systems with lifetimes as short as $0.1$ Gyr.
On the other hand, \cite{mat06} claimed that the fraction should be less than $\sim 35\%$ in order to explain the [O/Fe]-[Fe/H] relation and iron abundance in the intracluster medium.
In our model, although $50\%$ of SNe Ia eventually come from the MS+WD systems, the lifetime shorter than $0.1$ Gyr is possible only for such higher metallicity as $Z>Z_\odot$.
Figure \ref{fig:snlife} shows the time evolution of the cumulative function of SNe Ia lifetime in the solar neighborhood, where the star formation and chemical enrichment histories are taken into account.
At the galactic age of $t=3$ Gyr, only the MS+WD system can produce SNe Ia, and thus the lifetime of all SNe Ia is less than 1 Gyr.
The RG+WD systems contribute at later time.
At present ($t=13$ Gyr), the fraction of SNe Ia that have shorter lifetime than 1 Gyr is $\sim 50\%$.
The total fraction with $\le 0.1$ Gyr is only $10\%$ in our model.

\section{Galactic Supernovae Rate}

\begin{deluxetable}{lcc}
\tablenum{5}
\footnotesize
\tablewidth{0pt}
\tablecaption{\label{tab:galmodel}
The input parameters of galaxy models: timescales of the star formation and inflow in Gyr.}
\tablehead{
 & $\tau_{\rm s}$ [Gyr] & $\tau_{\rm i}$ [Gyr]
}
\startdata
E      & $0.1$  & $0.1$  \\
S0a/Sa & $1.8$  & $2.0$  \\
Sab/Sb & $1.75$  & $4.5$ \\
Sbc/Sc & $2.0$  & $8.3$  \\ 
Scd/Sd & $2.1$  & $23.0$ \\
\enddata
\end{deluxetable}

\subsection{SN Ia rate in Ellipticals}

The present SN Ia rate in elliptical galaxies gives also a stringent constraint on the SN Ia progenitor models.
Figure \ref{fig:ellsnr} shows the SN Ia rate history in ellipticals, which is similar to Fig.5 of K00, but for the cosmological parameters of $H_0=70$ km s$^{-1}$ Mpc$^{-1}$, $\Omega_0=0.3$, $\lambda_0=0.7$, the galactic wind epoch of $1.5$ Gyr, and the galactic age of $13$ Gyr.
(Note that the observed SN Ia rate is proportional to $H_0^2$.)
The star formation history is assumed as in K00;
the bulk of stars in ellipticals are formed at $z \gtsim 3$,
and thus having ages older than $10$ Gyr.
The adopted infall and star formation timescales are summarized in Table \ref{tab:galmodel}.
The present B-V color of $0.9$ \citep{rob94}, the mean stellar metallicity of $\sim -0.2$ \citep{kob99}, and the mean stellar [$\alpha$/Fe] of 0.2 \citep{spr09} are reproduced.
The present stellar mass-to-light ratio is $M/L_{\rm K}=1.8$ and $M/L_{\rm B}=9.3$ in our model.
The resultant SFR is similar to the SN II rate (black dot-dashed line).

\begin{itemize}
\item In the DD model (green short-dashed line),
the SN Ia lifetimes are too short to explain the observed SN Ia rate
at the present epoch \citep{cap99}.
Toward $z \sim 4$, the SN Ia rate rapidly increases, and the SN Ia rate at $z \sim 4$ is $10$ times larger than the present rate.

\item In the MR01-like model (blue long-dashed line), the fraction of SNe Ia with $t_{\rm Ia} \gtsim 10$ Gyr is larger than the DD model, so that the present rate is not too small compared with the observation.
Toward $z \sim 3$, the SN Ia rate gradually increases, and the SN Ia rate at $z \sim 1$ and $\sim 3$ is larger by a factor of $2$ and $4$, respectively, than the present rate.

\item 
In ellipticals, the chemical enrichment takes place so early that 
the metallicity effect on SNe Ia is not effective.
Thus the SN Ia rate depends almost only on the lifetime.
Our model (red solid line) includes the RG+WD systems with $t_{\rm Ia} \gtsim 10$ Gyr as the SN Ia progenitors, and thus the observed SN Ia rate in ellipticals can be well reproduced.
Toward higher redshift, the predicted SN Ia rate does not change very much (in the supernova units, SNu).

\item
In the K98 model (see Fig.5 in K00), the lifetime distribution shows a gap between the two components of the MS+WD and RG+WD systems (Fig. \ref{fig:life1}), so that two peaks are clearly seen in the SN Ia rate in ellipticals.
Such peaks are not clearly seen in our new SN Ia model because the lifetime ranges get gradually wider for higher metallicity (Fig. \ref{fig:life3}).
\end{itemize}

The sharp peak at $t \sim 1.5$ Gyr is caused both by the initial star burst assumed in the SFR and by the spikes seen in Figure \ref{fig:life3}, which are caused by the enhancement of the stellar lifetime from the occurrence of the He core flash.
The galactic winds might be driven by this enhancement of the SN Ia rate in elliptical galaxies. Therefore, we set the galactic wind epoch of $1.5$ Gyr in our model of elliptical galaxies.

\subsection{SN Ia rate depending on Galaxy Type}

Since the star formation history is different for different types of galaxies, the SN Ia and II rates depend on the morphological type of galaxies.
In the observations, the present rates are available for various types of galaxies, which can give a constraint on the SN Ia models.
In Fig.4 of K00, the observed ratios between SN Ia and II rates have been reproduced with our SN Ia model.
We show the similar figure to compare with the updated observation \citep{man05}.
We should note, however, that the absolute rates per B-band luminosity involves an uncertainty from dust extinction, and the rates per mass involve an uncertainty in the mass-to-light ratio.

We update the evolution models of four types of spiral galaxies, in order to meet the observational constraints on the present colors and the gas fractions, assuming the galactic age of 13 Gyr.
The adopted infall and star formation timescales are summarized in Table \ref{tab:galmodel}.
Figure \ref{fig:spisfr} shows the time evolution of (a) the SFR, (c) the gas fraction per luminosity, and (d) the B-V color for four types of spirals (red dotted line for S0a/Sa; yellow solid line for Sab/Sb; green long-dashed line for Sbc/Sc; blue short-dashed line for Scd/Sd).
In earlier-type of spirals, star formation takes place earlier, and thus the earlier-type of spirals have older ages of stellar populations, smaller gas fractions, and redder colors at present.
In the panel (b), the upper and lower four lines respectively show the SN II and Ia rates per B-band luminosity (in SNu). The observational data is taken from \cite{cap99} for S0a-Sb and Sbc-Sd.
The present SN II+Ibc rate is larger for later-type of spirals, while the SN Ia rate in SNu is not different so much among the various types of galaxies, which are roughly consistent with the observation.

Figure \ref{fig:snrgal} shows the present supernova rates per K-band luminosity (upper panel) and per mass (lower panel) against the morphological type of galaxies.
The supernova rates are larger for later-type of spirals because the SFR is higher.
Almost parallel relations between the observed rate and the galaxy type are seen for SNe II and Ibc, which may suggest the binary fraction is universal being independent of the galaxy type.
Among core-collapse supernovae, we assume that 85\% and 15\% are SNe II (blue dashed line) and SNe Ibc (green dotted line), respectively.
Thus, it seems to be reasonable to adopt the same $b$ parameters as SNe Ia for all types of galaxies.
As a result, our SN Ia model (red solid line) gives excellent agreement with the observed SN Ia rates.
In our SN Ia models, there are two types of progenitor systems;
The SN Ia rate is larger for later-type spirals, which is due to the young population of the MS+WD systems.
The slope of the SN Ia rate against the galaxy type is flatter than that of the SN II and Ibc rates, which is due to the old population of the RG+WD systems.

The observational data is taken from \cite{man05}, where the adopted mass-to-light ratios are $M/L_{\rm K} \sim 0.8, 0.7, 0.5$, and $0.4$ for E/S0, S0a/b, Sbc/d, and Irr in their Table 2.
In our models, the ``stellar'' mass-to-light ratios are $M/L_{\rm K} \sim 1.29, 1.14, 1.07$, and $1.03$ for S0a/Sa, Sab/Sb, Sbc/Sc, and Scd/Sd, which are $\sim 2$ times larger than their ratios.
This difference might come from the difference in the IMF.
We thus plot the observational data multiplied by a factor of $2$ for the rates per mass.

Figure \ref{fig:snrgal-all} compares the galactic supernova rates for different SN Ia models.
\begin{itemize}
\item In the DD model (green short-dashed line), the lifetime of the majority of SNe Ia is short, no large difference between the SN II and Ia rates is seen in the trend against the galaxy type, which is not consistent with the observations.
\item The MR01-like model (blue long-dashed line) and our SD models (red solid line and cyan dotted line) give almost the same trends, regardless of the metallicity effect, as far as the lifetime distributions in this work and K98 are applied.
\end{itemize}

\subsection{Cosmic Supernovae Rate}

Galaxies have various
timescales for star formation and chemical enrichment, and the occurrence of
SNe Ia depends on the metallicity therein.
Therefore, we should calculate the cosmic supernova rate by summing up the supernova 
rates in spirals and ellipticals with the ratios of the relative mass
contribution in the Universe (see K00 for the detail).
As in Fig.6 of K00, the resultant cosmic SFRs shows a peak at $z \sim 2$.
Figure \ref{fig:snr} shows the cosmic supernova
rates with our new SN Ia model (solid lines), 
as a composite of those in spirals (blue lines) and
ellipticals (red lines)
in SNu (upper panel) 
and in the unit of the comoving density (lower panel), respectively.
The SN II rate (green dot-dashed lines) traces the star formation history, and shows a peak at $z \sim 2$ in the comoving density, being consistent with the observations \citep{cap05,bot07}.

As in spirals,
the total SN Ia rate (green solid lines) gradually decreases toward $z \sim 2.6$ in the unit per luminosity,
while the total SN Ia rate slightly increases to $z \sim 1$ and then decreases in the unit per volume.
In ellipticals, the chemical enrichment takes place so quickly that 
the metallicity is large enough to produce SNe Ia at $z \gtsim 2$.
These are broadly consistent with the observations (e.g., 
\citealt[Supernova Cosmology Project, 38 SNe Ia]{pai02};
\citealt[High-z Supernova Search Team, 8 SNe Ia]{ton03};
\citealt[Great Observatories Origins Deep Survey, 25 SNe Ia]{dah04};
\citealt[Institute for Astronomy Deep Survey, 98 SNe Ia?]{bar06};
\citealt[Supernova Legacy Survey, 58 SNe Ia]{nei06}).
Around $z \sim 0.5$,
the total SN Ia rate may be smaller than the observations,
although \citet{poz07} shows a significant lower rate with the Subaru Deep Field.

The cosmic SN Ia rate history for different SN Ia models are also shown in Figures \ref{fig:snr} and \ref{fig:snr2}.
\begin{itemize}
\item If we do not include the metallicity effect (dotted lines in Fig. \ref{fig:snr}), the SN Ia rate in spirals does not decrease so much, and the total SN Ia rate per luminosity is almost constant. The rate per volume slightly increases toward higher redshifts, and is larger than the present rate by a factor of $\sim2$ at $z\sim2$.
In the observations, the decrease in the total rate is seen, which supports the existence of the metallicity effect.
\item In the DD (short-dashed lines in Fig. \ref{fig:snr2}) and MR01-like (long-dashed lines in Fig. \ref{fig:snr2}) models, the lifetime of the majority of SNe Ia is so short that the rate in ellipticals keeps on increasing at $z\ltsim 4$, and that the rate in spirals shows a peak at $z \sim 1.5$.
As a result, the total SN Ia rate per luminosity is almost constant, and the rate per volume keeps on increasing toward higher redshifts.
The total SN Ia rate is larger than the present rate by a factor of $\sim2$ and $\sim4$ at $z\sim1$ and $2$, respectively.
\item In the K98 model (see Fig. 6 in K00), two peaks are seen in the SN Ia rate history in ellipticals, which are not clearly seen in our new SN Ia model, because the lifetime ranges get gradually wider for higher metallicity (Fig. \ref{fig:life3}).
\end{itemize}

The observed cosmic SFR shows an increase by a factor of $\sim 10$ from $z=0$ to $\sim 1$ \citep[e.g., ][]{sch05}, and such a large increase is not seen in the composite SFRs of these models for ellipticals and spirals.
Such a large increase in the SFR results in the rapid increase in SN II and Ia rates from $z=0$ to $\sim 1$ that may be seen in the observed SN Ia rates.
By modifying the SFRs in spirals as well as including the contribution of irregular galaxies, there can be a better fit \citep{cal06}.
However, we don't apply such parameter-tuning in this paper because other effects are more important in the cosmic chemical enrichment: i) the number evolution of each type of galaxies, ii) merging of galaxies, iii) internal structure, namely, metallicity gradients within galaxies, iv) the contribution of dwarf galaxies.
We will rather show the cosmic supernova rates with fully self-consistent cosmological simulations \citep{kob07,kob08}.
Nonetheless, our result with the simple assumptions is robust; the SN Ia rate is high in ellipticals and drops in spirals at high-redshift.

\section{Discussion}

\subsection{SN Ia Formulation}

In galactic chemical evolution models, the following two formulations have been proposed. Their parameterizations are very different, and it would be worth summering here. Observations of binary systems could give a clue to improve these formulations.

(1) The formulation of the SN Ia rate has been first proposed by Greggio \& Renzini (1983).
The SN Ia rate is described as
\begin{equation}
{\cal R}_{\rm Ia}=A~
\int_{M_{\rm B,inf}}^{M_{\rm B,sup}}\,
\phi(M_B)
\int_{\mu_{\rm min}}^{\mu_{\rm max}}\,
f(\mu) \, \psi(t-\tau_{M_2}) \,d\mu\,dM_{\rm B} .
\label{eqmat}
\end{equation}
The first integral is for the total mass of the binary $M_{\rm B}=M_1+M_2$, and the second integral is for the mass fraction of the secondary star, $\mu=M_2/M_B$.
In MR01, the Scalo IMF $\phi$ is adopted, and the distribution function of the mass fraction is described as $f(\mu)=2^{1+\gamma}(1+\gamma)\mu^\gamma$. 
$\gamma=2$ is adopted in their successful models, which gives weight for massive secondaries in the distribution function of mass ratios (see Fig. \ref{fig:binary}).
The upper and lower limits of integrals are $M_{\rm B,inf}=\max[2M_2(t),M_{\rm Bm}]$ and $M_{\rm B,sup}=M_{\rm BM}/2+M_2(t)$, where the limits to the total mass of the binary are $M_{\rm Bm}=3M_\odot$ and $M_{\rm BM}=16M_\odot$. 
Therefore, the initial mass range of the primary stars is $1.5 M_\odot \leq M_1 \leq 8 M_\odot$ and the secondary stars are AGB stars with the initial mass of $M_2 \leq 8 M_\odot$.
The lower limit of the primary, however, should not $1.5 M_\odot$ but $\sim 3 M_\odot$, since the primary should be a C+O WD.
Although $A$ is called as the binary fraction, $A$ actually indicates the fraction of binary systems with suitable separation to become SNe Ia \citep{gre83}.

(2) Based on the SD scenario, an alternative formulation has been proposed by K98 and K00. The SN Ia rate is given as
\begin{equation}
{\cal R}_{\rm Ia}=b~
\int_{\max[m_{{\rm p},\ell},\,m_t]}^{m_{{\rm p},u}}\,
\frac{1}{m}\,\phi(m)~dm~
\int_{\max[m_{{\rm d},\ell},\,m_t]}^{m_{{\rm d},u}}\,
\frac{1}{m}\,\psi(t-\tau_m)\,\phi_{\rm d}(m)~dm  .
\label{eqkob}
\end{equation} 
These integrals are calculated separately for the primary and secondary stars ($m_{\rm p}$ and $m_{\rm d}$), respectively.
In our works, the Salpeter IMF $\phi$ is adopted, and the distribution function of the companion's mass is assumed to be the power-law as 
$\phi_{\rm d}(m)\propto m^{-0.35}$.
The slope $-0.35$ is chosen to meet the observations \citep[e.g.,][]{duq91}, and such negative slope gives the distribution function weighted for less-massive secondaries (see Fig. \ref{fig:binary}).
$\phi_{\rm d}$ is normalized to unity in the integrated mass ranges as $\int_{m_{{\rm d},\ell}}^{m_{{\rm d},u}}\,\frac{1}{m}\,\phi_{\rm d}(m)\,dm=1$.
Since the primary stars are C+O WDs, the initial mass range of the primary stars is $m_{\rm p} = 3-8 M_\odot$. 
The initial mass ranges of the companion stars, $m_{{\rm d},\ell}$ and $m_{{\rm d},u}$, are given by the simulation of binary evolution (Hachisu et al. 1999ab) for the RG+WD and MS+WD systems.
The binary parameter $b$ denotes the 
total fraction of primary stars that eventually explode as SNe Ia,
and is determined from the chemical evolution in the solar neighborhood.
$b$ includes the binary fraction and the efficiency for each binary system.
In addition, the metallicity effect is taken into account (\S 2).
The lowest metallicity to produce SNe Ia was set to be [Fe/H]$= -1.1$
both in K98 and K00\footnote{The note in MR01 is incorrect. See K98, $\ell$.16 in the right column, p.156, and K00, $\ell$.15 in the right column, p.27. }.
In our new model (\S 2.2), the mass stripping effect is taken into account, and $m_{{\rm d},\ell}$ and $m_{{\rm d},u}$ are given as a function of metallicity as in Table \ref{tab:sniaparam}.

Strictly speaking, $m_{{\rm d},\ell}$ and $m_{{\rm d},u}$ depend on $m_1$, but it can be neglected.
For $M_{\rm WD}=0.7 M_\odot$, the mass range is a little narrower, and thus the
shortest lifetime is 0.2 Gyr that is longer than 0.1 Gyr for $M_{\rm WD}=1.0
M_\odot$ ($Z=0.02$). 
The range of the orbital period is much narrower, so
that the contribution of $M_{\rm WD} < 1.0 M_\odot$ is smaller. On the other
hand, for $M_{\rm WD}=1.1 M_\odot$, the shortest lifetime is a little shorter,
but the number of such WD supposed to be very small. As a result, the
[X/Fe]-[Fe/H] relation is mainly determined by the contribution for
$M_{\rm WD}=1.0 M_\odot$.

The results slightly depend to some extent on the stellar lifetime function $\tau_m$ of low and intermediate mass stars.
Figure \ref{fig:lifetime} shows the adopted stellar lifetime function as a function of initial mass and metallicity, which is provided by Kodama \& Arimoto (1997).
The stellar evolution tracks are calculated with the code described in Iwamoto \& Saio (1999) with the core overshooting.
Because of the coarse mass grid, there is a tiny irregularity at $1.8 M_\odot$ ($Z=0.004, 0.008$), $2.0 M_\odot$ ($Z=0.02$), $2.2 M_\odot$ ($Z=0.0001$-$0.002$), and $3.3 M_\odot$ ($Z=0.0002$) in the table, which we have smoothed with interpolation.
The stellar lifetime function overall agrees with more recent models such as \citet{kar07}.
The long and short dashed-lines are for the analytical functions adopted in MR01 and HKN08, respectively, where the metallicity effect is neglected.
With those analytical functions, the stellar lifetime is systematically shorter.

We adopt the Salpeter IMF with a slope of $-1.35$.
\citet{kro07} showed a similar slope ($-1.3\pm0.5$) for $m \ge 0.5 M_\odot$, which is flatter than in the Scalo IMF.
For $m < 0.5 M_\odot$, the slope is shallower and the number of brown dwarfs are smaller than in the Salpeter IMF.
In chemical evolution models, however, metal enrichment is proceeded only by stars with $m \ge 0.5 M_\odot$.
If a suitable value is chosen for the lower mass-cut of the IMF, the Salpeter IMF can give equal results with the Kroupa IMF.
Therefore, we adopt the single slope for a mass range of $m=0.07-50 M_\odot$.
This does not necessarily mean that there is no star at $m < 0.07 M_\odot$. The requirement is that the mass fraction of $m=0.5-50 M_\odot$ is 45\%.

\subsection{Binary Parameters}

In the two formulations, Eqs.[\ref{eqmat}] and [\ref{eqkob}], the treatment of binary systems is different for the following three points: i) binary fraction, ii) mass ratio, and iii) binary IMF.
These differences may reflect the formation processes of binary systems. Theoretically, the prompt fragmentation scenario seem to be more feasible than captures or delayed breakup of accretion disks \citep{toh02}, and these scenarios will be tested with the observed distribution functions of the binary parameters.
In chemical evolution models, however, these uncertainties can be cancelled out by the choice of $b$ or $A$.

i) {\it Binary fraction ---}
The total number of SNe Ia, namely, the binary fraction $A$ or binary parameters $b$, could be determined from the binary population synthesis.
However, the evolution of binary systems is complicated and the effect of the WD winds is not included.
In Eq.[\ref{eqkob}], we treat $b$ as parameters being determined for the RG+WD and MS+WD systems, independently, from the requirement to meet the observed MDF of G-dwarf stars and the [O/Fe]-[Fe/H] relation in the solar neighborhood.
In K98, we adopted $b_{\rm RG}=b_{\rm MS}=0.04$
from the Kolmogorov-Smirnov (KS) test of the [O/Fe]-[Fe/H] relation, but
in K00, we adopted $[b_{\rm RG}, b_{\rm MS}]=[0.02, 0.05]$
from the $\chi^2$ test\footnote{There are two miss-prints in Appendix of K00, not $[b_{\rm MS}=0.02, b_{\rm RG}=0.05]$ but $[b_{\rm RG}=0.02, b_{\rm MS}=0.05]$.}.
In K06, we adopted smaller values of [0.02, 0.04] because Fe are also produced by HNe in the updated nucleosynthesis yields.
For our new SN Ia model, $[b_{\rm RG}, b_{\rm MS}]=[0.023, 0.023]$ are chosen at $Z=0.004$, which becomes [0.040, 0.032] at $Z=0.05$, as discussed in \S 3.
In total, the fraction of SNe Ia is $5-7$\%, which is consistent with the constraints in \citet{mao08} except for the X-ray observations.

In Eq.[\ref{eqmat}], the definition of $A$ is totally different from that of $b$, and $A$ can be simply determined from the MDF.
This is misunderstood in Matteucci \& Recchi (2001), where they claimed that they failed to reproduce the K98 results (Mod.1 in MR01).
We would demonstrate in Figure \ref{fig:ofe1} that we can reproduce their and our results.
For the comparison, we adopt the same IMF, stellar lifetimes, and nucleosynthesis yields as in our models.
Adopting MR01's formula Eq.[\ref{eqmat}] and parameters, we almost reproduce Mod.1 and Mod.3 in MR01 (short- and long-dashed lines, respectively).
Here we adopt $A=0.035$ to meet the MDF, which is better than $A=0.05$ originally adopted in MR01.

In Mod.3 in MR01, $A=0.02$ and $0.05$ are adopted for the RG+WD and MS+WD systems, respectively, which are the same values of the binary parameter $b$ in K00.
However, since $\phi_{\rm d}$ is normalized to unity not for $m = 0.07-50M_\odot$ but for $m_{\rm d} \sim 1-3M_\odot$, the value of $A$ should be ten times larger than the value of $b$.
Adopting $A=0.45$,
we reproduce our original K98 model (solid line) even with the MR01 formula (dotted line).
With the MR01 formula,
the slope of [O/Fe] is slightly different, and [O/Fe] around [Fe/H] $\sim -0.5$ is slightly higher than the K98 model.
The best choice of $A$ depends on the two other assumptions (i and ii), and varies in the range of $A=0.10-0.45$ as shown below.
This value of $A$ is larger than in the MR01-like model because the adopted mass ranges both for primary and secondary stars are narrower.

ii) {\it Mass ratio ---}
In Figure \ref{fig:binary}, we show several functions of the mass ratio $f(q)$ or the mass fraction $f(\mu)$ (Note that $q\equiv m_2/m_1=\frac{\mu}{1-\mu}$ and $\frac{d\mu}{dq}=\frac{1}{(1+q)^2}=(1-\mu)^2$).
The function that is introduced in MR01 is weighted for massive secondaries (i.e., the positive value of $\gamma$, long-dashed line).
Among the binary population synthesis, several functions are adopted: 
$f(q)\propto (1+q)^{-2}$, i.e., constant $f(\mu)$ (dotted line, \citealt{por95}), and
constant $f(q)$, i.e., $f(\mu)\propto 1/(1-\mu)^2$ (short-dashed line, \citealt{han95}).
Observations of binary systems, however, suggest that the function is weighted for less-massive secondaries, as in our function (solid line).
The histogram shows the observation of G-dwarfs by \citet{duq91}.
For B-type stars \citep{sha02,kou05}, the power-law distributions with the slopes of $-0.33$ to $-0.5$ are reported, and the random paring of stars is not supported.
Note that the shape of $f(q)$ is affected by selection bias \citep{hog92}, and may depend on the density and composition of the stellar aggregate where the binaries have been formed \citep{kro95}.
Our function is between these observations;
the secondary IMF $\phi_{\rm d}$ is weighted for less-massive secondaries, and the slope is shallower than that of the primary IMF $\phi$.

In Figure \ref{fig:ofe2}, we show that the mass ratio function does not change our results.
In other words, it is possible to obtain almost identical results with different sets of $A$ and $f(\mu)$.
For our SN Ia model, although MR01 adopted $\gamma=-0.35$, $\gamma=-1.35$ is correct because the number function should be the mass function $\phi_{\rm d}$ multiplied by $1/m$.
Since $f(\mu)$ with $\gamma=-1.35$ is weighted for low-mass secondaries, the relative contribution of the MS+WD systems is small, and $A$ should be as large as $0.45$.
With the constant $f(\mu)$ or constant $f(q)$, the contribution of the MS+WD systems is as large as that of the RG+WD systems, and thus $A$ becomes $0.30$ and $0.25$, respectively.
With $\gamma=2$ as in MR01, $A=0.20$ gives the best fit.

iii) {\it Binary IMF ---}
The IMF is applied to the total mass of the binary in Eq.[\ref{eqmat}], while it is applied to the mass of the primary star in Eq.[\ref{eqkob}].
This may correspond to different approaches in the formation of binaries: a common or individual formation.
This treatment also does not change our results.
If we adopt not $\phi(m_{\rm B})$ but $\phi(m_1)$, the normalization is changed by a factor. For $\gamma=-1.35$, constant $f(\mu)$, constant $f(q)$, and $\gamma=2$, $A$ becomes $0.25$, $0.15$, $0.12$, and $0.10$, respectively.

Needless to say, for the models that give the same [O/Fe]-[Fe/H] relation, the SN Ia rate history is almost the same.
In the lower panels of Figures \ref{fig:ofe1} and \ref{fig:ofe2}, the time evolution of SN Ia rate is shown.
Only 0.05 dex difference is seen in the present SN Ia rate among the models with different binary mass fractions.
The present SN Ia rate with $\gamma=-1.35$ is 0.1 dex larger than with $\gamma=2$.
It seems not possible to put constraint on the binary mass fraction and binary IMF from the observed SN Ia rate because of the large errorbar.

As a summary, once a suitable $b$ or $A$ is adopted, almost identical results can be produced with either of these two formulations.
The most important factor is the mass ranges of companion stars, which changes the distribution function of SN Ia lifetimes.
The total number of SNe Ia, $b$ or $A$, are determined from the observational constraints in this paper, but may be larger than what can be obtained with binary population synthesis models \citep[e.g.,][]{mao08}.
Other assumptions, the mass ratio function and IMF, are not very important since these two effects can be cancelled out by the choice of the total number of SNe Ia.
Observations of binary systems and the theory of binary formation could give a clue to solve this degeneracy, and may put more constraints on SN Ia models.

\begin{deluxetable}{lcccc}
\tablenum{6}
\footnotesize
\tablewidth{0pt}
\tablecaption{\label{tab:summary}
Test results and predictions of the SN Ia models.}
\tablehead{
\colhead{} &
\multicolumn{2}{c}{SD} &
\colhead{MR01} &
\colhead{DD} \\
\colhead{metallicity effect} &
\colhead{yes} &
\colhead{no} &
\colhead{} &
\colhead{}
}
\startdata
Lifetime [Gyr] & \multicolumn{2}{c}{$0.1-20$ peak at $0.1$, $1$} & $\sim 0.3$ & $\sim 0.1$ \\
${\rm [}\alpha$/Fe]-[Fe/H] relation & $\bigcirc$ & $\times$ & $\times$ & $\times$ \\
Present SN Ia rate in Es & $\bigcirc$ & $\bigcirc$ & $\bigcirc$ & $\times$ \\
Cosmic SN Ia rate history & peak at $z\sim1$ & constant & increase & increase \\
SNe Ia observed at $z\gtsim 2$ in Ss? & few & yes & yes & yes \\
\enddata
\end{deluxetable}

\section{Conclusions}

\begin{figure}
\center
\includegraphics[width=18cm]{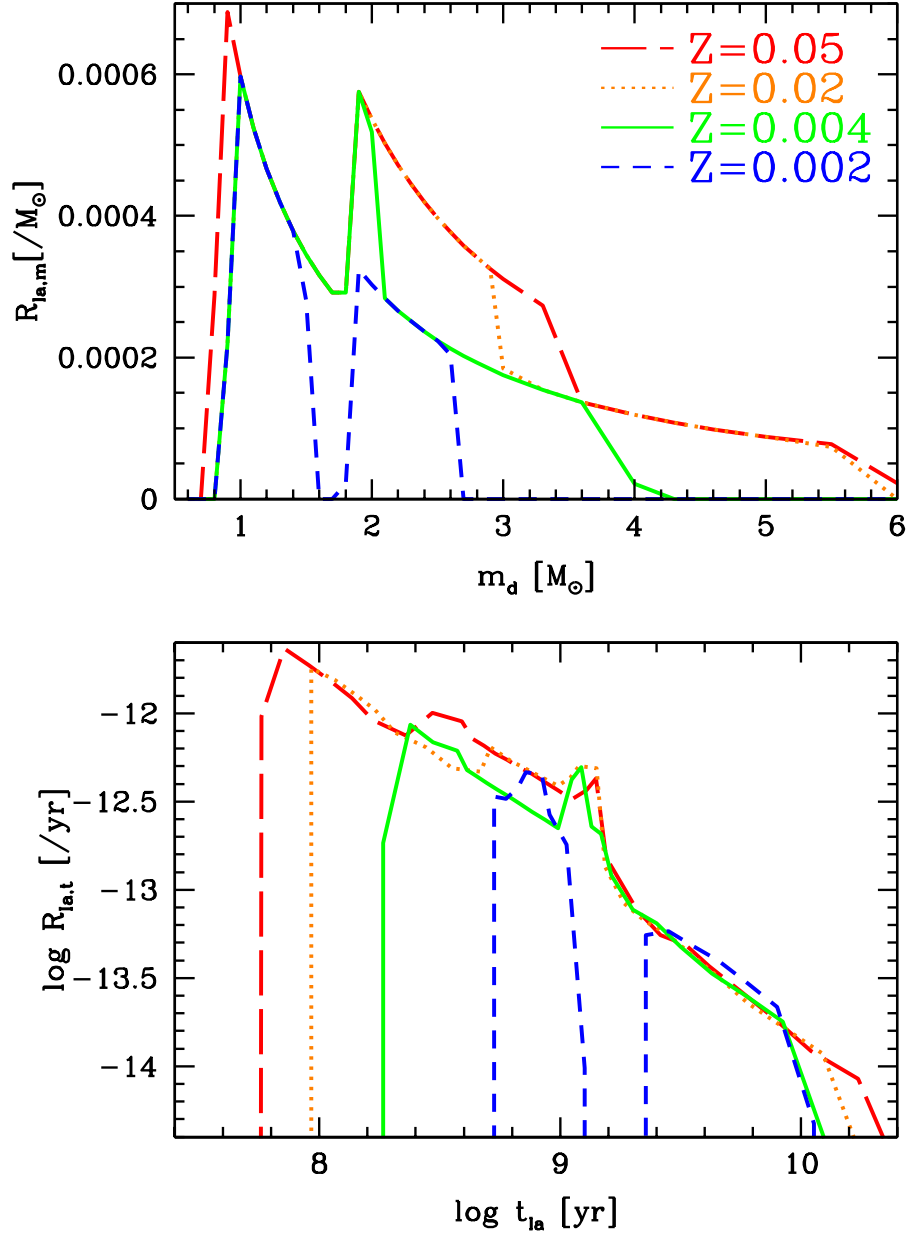}
\caption{\label{fig:life3}
The distribution functions of the progenitor mass (upper panel) and lifetime (lower panel), i.e., the SN Ia rates of stars with the same ages, for our new models with the metallicity $Z=0.002$ (blue short-dashed line), $Z=0.004$ (green solid line), $Z=0.02$ (orange dotted line), and $Z=0.05$ (red long-dashed line).}
\end{figure}

\begin{figure}
\center
\includegraphics[width=17cm]{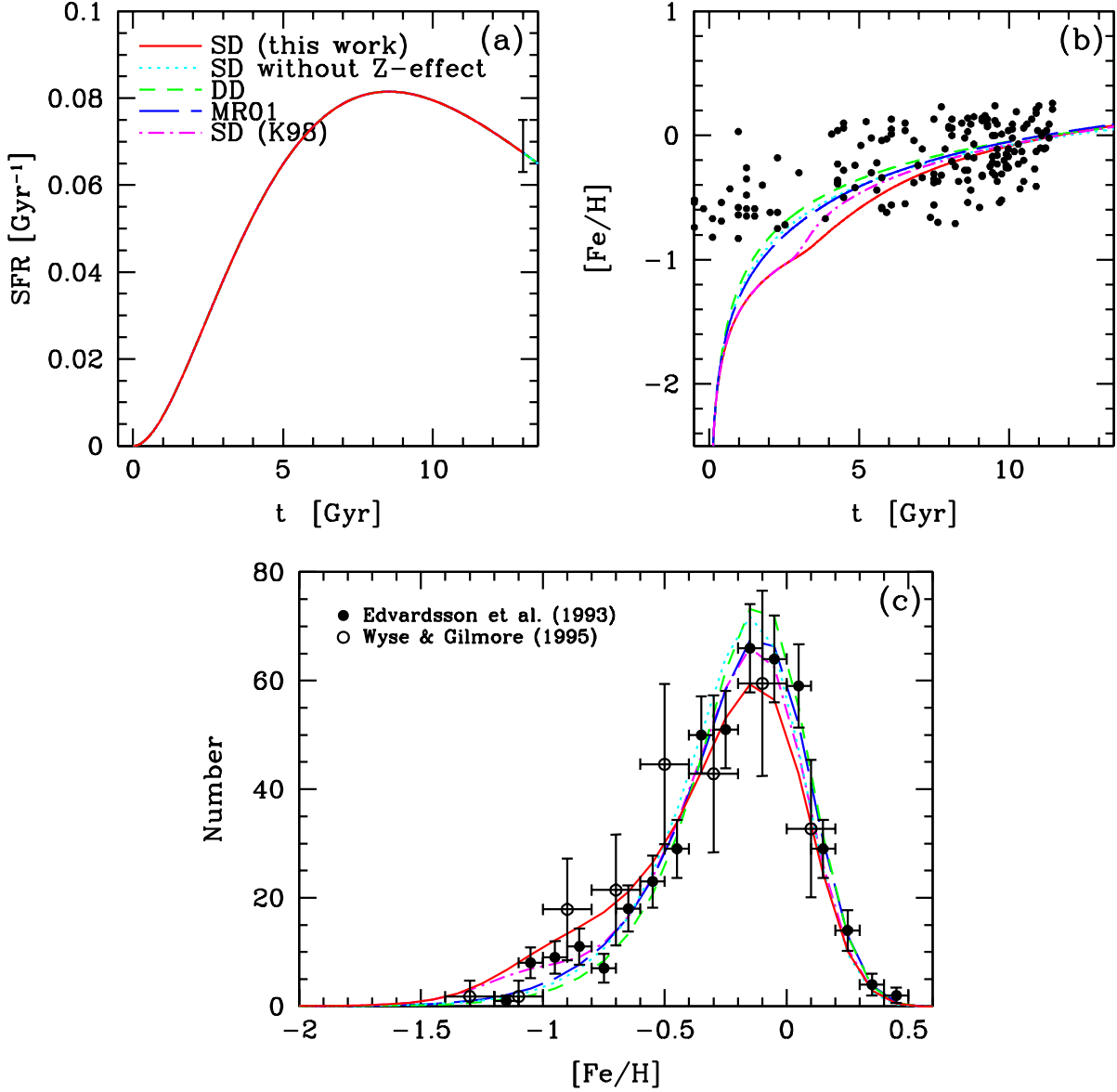}
\caption{\label{fig:mdf}
The chemical evolution of the solar neighborhood: (a) the star formation rate, (b) the age-metallicity relation, and (c) the metallicity distribution function
for different SN Ia models:
our SD model (red solid line), SD model without metallicity effect (cyan dotted line), DD model (green short-dashed line), MR01 model (blue long-dashed line), and K98 model (magenta dot-dashed line).
Observational data sources are:
\citet{mat97}, errorbar in panel (a);
\citet{edv93}, filled circles in panels (b) and (c); 
\citet{wys95}, open circles in panel (c).
}
\end{figure}

\begin{figure}
\center
\includegraphics[width=16cm]{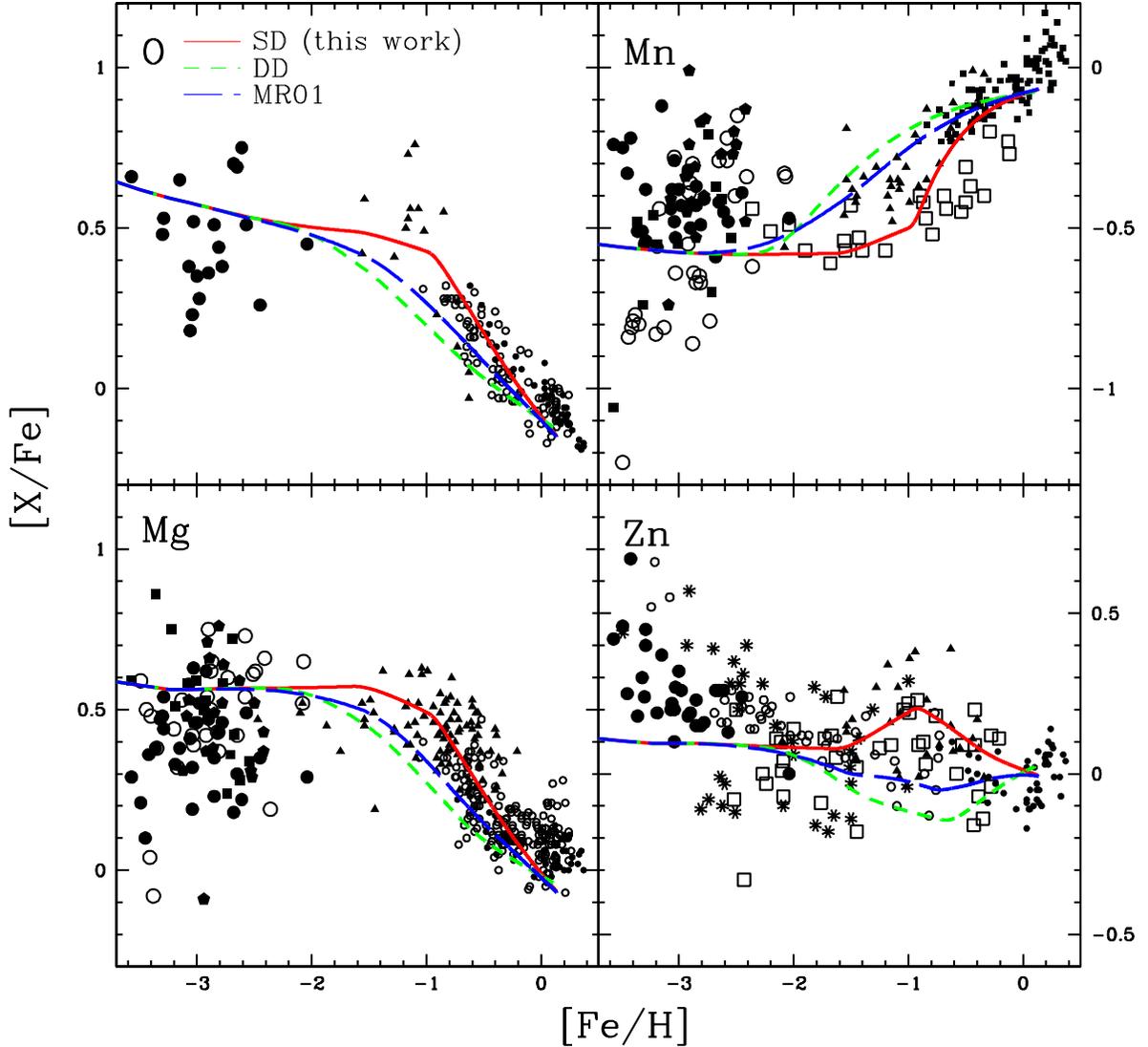}
\caption{\label{fig:xfe}
The evolution of the elemental abundance ratios, [X/Fe]-[Fe/H] relations, in the solar neighborhood
for different SN Ia models:
our SD model (red solid line), DD model (green short-dashed line), and MR01 model (blue long-dashed line).
Observational data sources are:
For disk stars, \citet{edv93}, small open circles; 
thin disk stars in \citet[2004]{ben03}, small filled circles; 
accretion component in \citet{gra03}, triangles.
For halo stars, \citet{mcw95}, large open circles;
\citet{rya96}, filled squares;
\citet{cay04}, large filled circles;
\citet{hon04}, filled pentagons.
For Mn, \citet{gra89}, open squares;
\citet{fel07}, small filled squares.
For Zn, \citet{sne91}, open squares;
\citet{pri00}, eight-pointed asterisks;
\citet[2007]{nis04}, open circles.
}
\end{figure}

\begin{figure}
\center
\includegraphics[width=16cm]{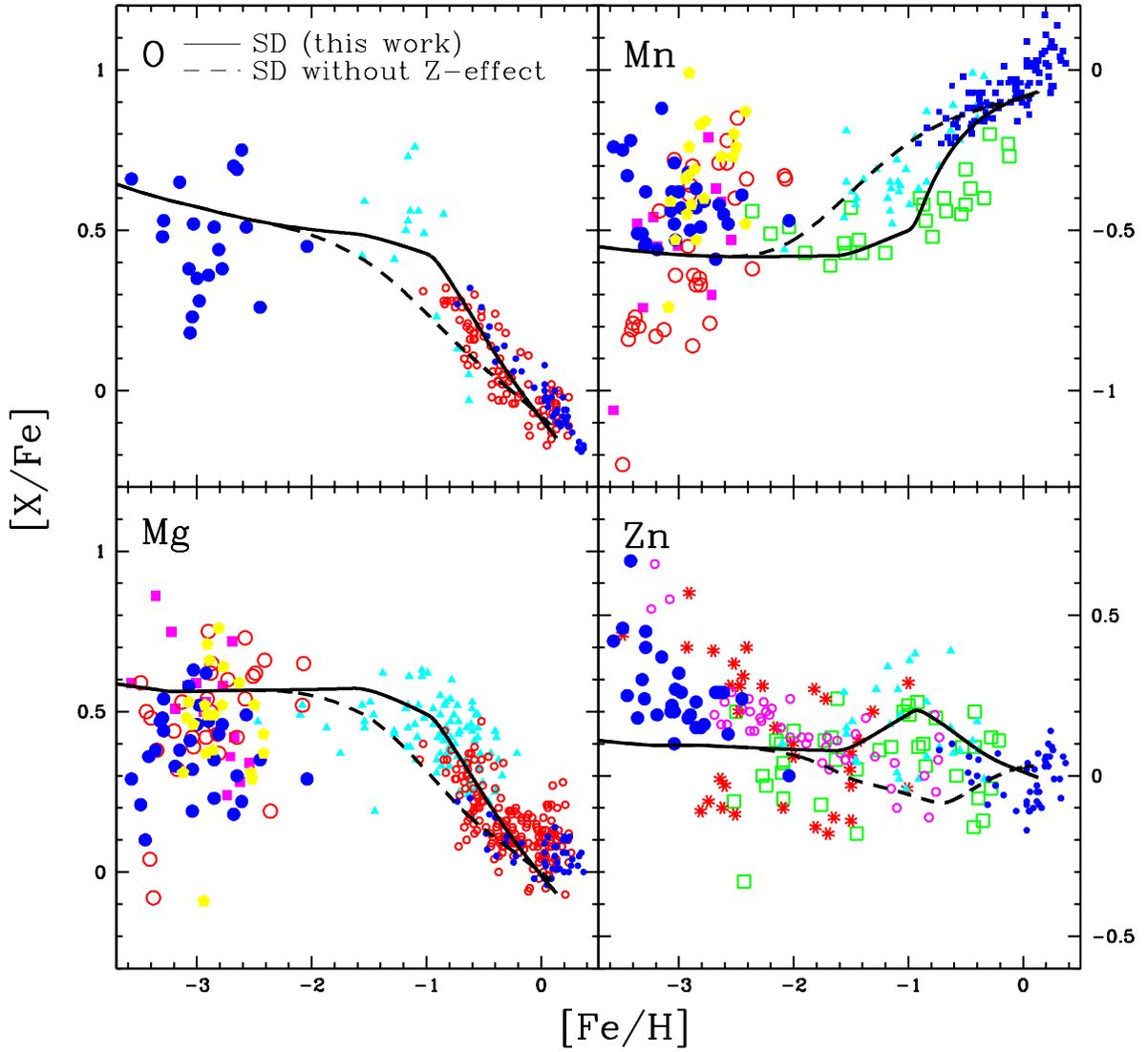}
\caption{\label{fig:xfe2}
Same as Figure \ref{fig:xfe} but for our SD model with (solid line) and without (dotted line) metallicity effect.
}
\end{figure}

\begin{figure}
\center
\includegraphics[width=16cm]{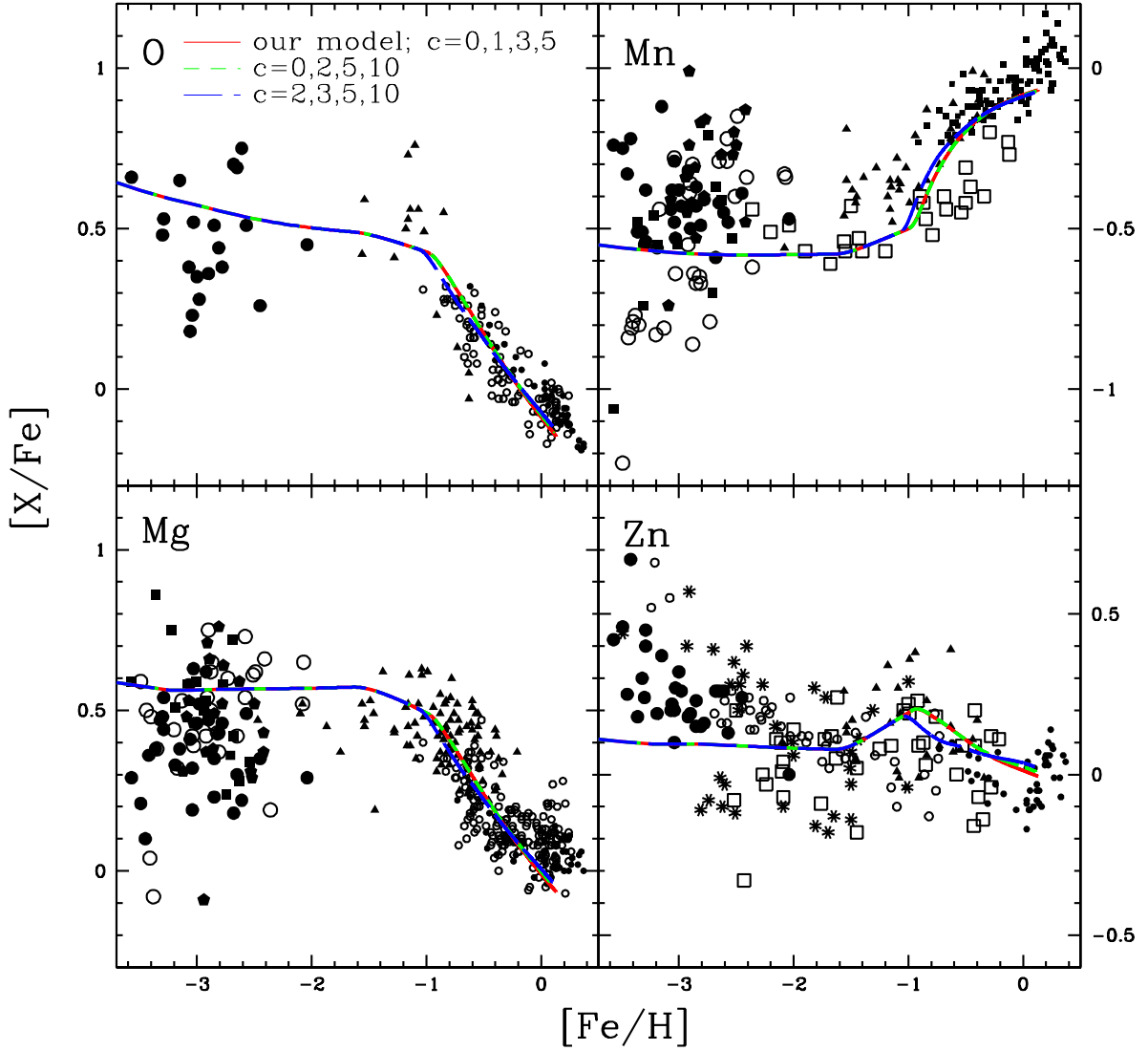}
\caption{\label{fig:xfe-c}
Same as Figure \ref{fig:xfe} but with different $c$; $c=(0,1,3,5)$ (solid line, main model), $c=(0,2,5,10)$ (short dashed line), and $c=(2,3,5,10)$ (long-dashed line) respectively for $Z=0.002, 0.004, 0.02, 0.05$.
}
\end{figure}

\begin{figure}
\center
\includegraphics[width=16cm]{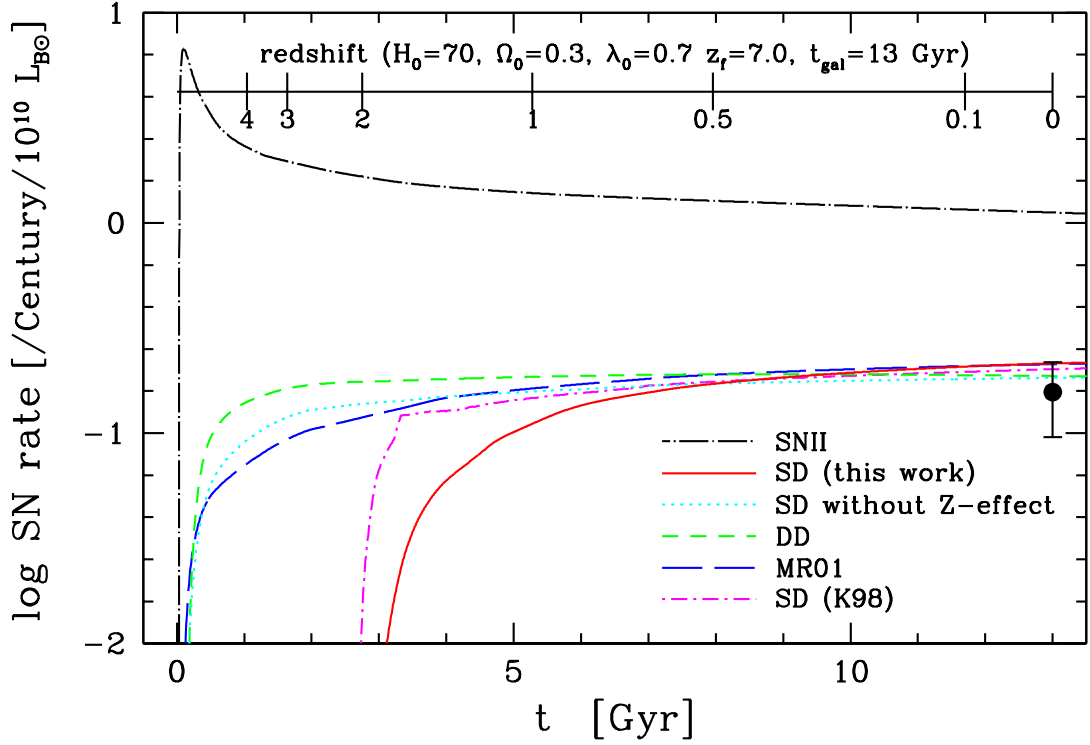}
\caption{\label{fig:snsnr}
Same as Figure \ref{fig:mdf} but for the SN Ia rate history in the solar neighborhood
with different SN Ia models:
our SD model (red solid line), SD model without metallicity effect (cyan dotted line), DD model (green short-dashed line), MR01 model (blue long-dashed line), and K98 model (magenta dot-dashed line).
The dot-dashed line shows the SN II rate.
The dot shows the observed SN Ia rate in S0a-Sb type galaxies \citep{cap99}.
}
\end{figure}

We construct a new SN Ia model, based on the SD scenario for the SN Ia progenitors, taking account of the metallicity dependences of the WD wind (K98) and the mass-stripping effect on the binary companion star (HKN08).
The lifetime distribution of SNe Ia is determined from the main-sequence mass range of the companion stars in the WD binary systems, and is given as a function of metallicity (Fig. \ref{fig:life3}).
The SN Ia rate in the systems with [Fe/H] $\ltsim -1$, e.g., $r \gtsim 10r_{\rm e}$ in intermediate-mass galaxies, $r \gtsim 50$ kpc in dwarf galaxies, and high-z spiral galaxies, is supposed to be very small (\S \ref{sec:zeffect}).

Our model naturally predicts that the SN Ia lifetime distribution spans a range of $t_{\rm Ia} \sim 0.1-20$ Gyr with the double peaks at $\sim 0.1$ and $1$ Gyr reflecting the two systems of the companion stars: the MS+WD and RG+WD systems, respectively.
With the chemical enrichment history in the Galactic disk, the fraction of the young population of SNe Ia is $10\%$ and $50\%$ for $t_{\rm Ia}<0.1$ and $1$ Gyr, respectively (Fig. \ref{fig:snlife}).
In the galaxies with a initial star burst, the SN Ia rate shows a strong peak at $t \sim 1.5$ Gyr, which may generate the SN Ia driven galactic winds.

We make a comparison of the lifetime distribution functions derived from different SN Ia models, and evaluate the results from the observational constraints as summarized in Table \ref{tab:summary}.

\noindent
(1) Contrary to the argument in Matteucci \& Recchi (2001, MR01), our SN Ia models give better reproduction of the [$\alpha$/Fe]-[Fe/H] relation in the solar neighborhood.
In the DD and MR01-like models, the typical lifetimes of SNe Ia are $\sim 0.1$ and $0.3$ Gyr, respectively, which results in the too early decrease in [$\alpha$/Fe] at [Fe/H] $\sim -2$.
If we do not include the metallicity effect in our model, [$\alpha$/Fe] decreases too early because of the shortest lifetime, $\sim 0.1$ Gyr, of the MS+WD systems.
In other words, the metallicity effect is more strongly required in the presence of the young population of SNe Ia, to be consistent with the chemical evolution of the solar neighborhood.

\noindent
(2) At the same [Fe/H], [Mn/Fe] starts to increase toward higher metallicity, because SNe Ia produce [Mn/Fe] $>0$. This evolutionary change is started from the same [Fe/H] as the decreasing trends of [$\alpha$/Fe].
[Zn/Fe] evolves differently for the different SNe Ia models, because of the combination of the different lifetime of SNe Ia and the metallicity effect on Zn yields of SNe II.
Further observations of these elements at [Fe/H] $\gtsim -1.5$ is critically important in order to identify the SN Ia progenitors and to clarify the metallicity dependence on the nucleosynthesis yields.
The lack of the similar [Mn/Fe] increase in dSphs suggests that those are not enriched by SNe Ia.

\noindent
(3) To explain the observed SN Ia in ellipticals at $z=0$, the lifetimes of a sufficient fraction of SNe Ia should be longer than $\sim 10$ Gyr, which is satisfied in our SD model and the MR01-like model, but not in the DD model.
Our SN Ia models are also successful in reproducing the SN II, Ibc, and Ia rates with their dependence on the galaxy type \citep{man06};
This cannot be reproduced with the DD model.
The large SN Ia rate in radio galaxies could be explained with the young population of the MS+WD systems in our model.

\begin{figure}
\center
\includegraphics[width=16cm]{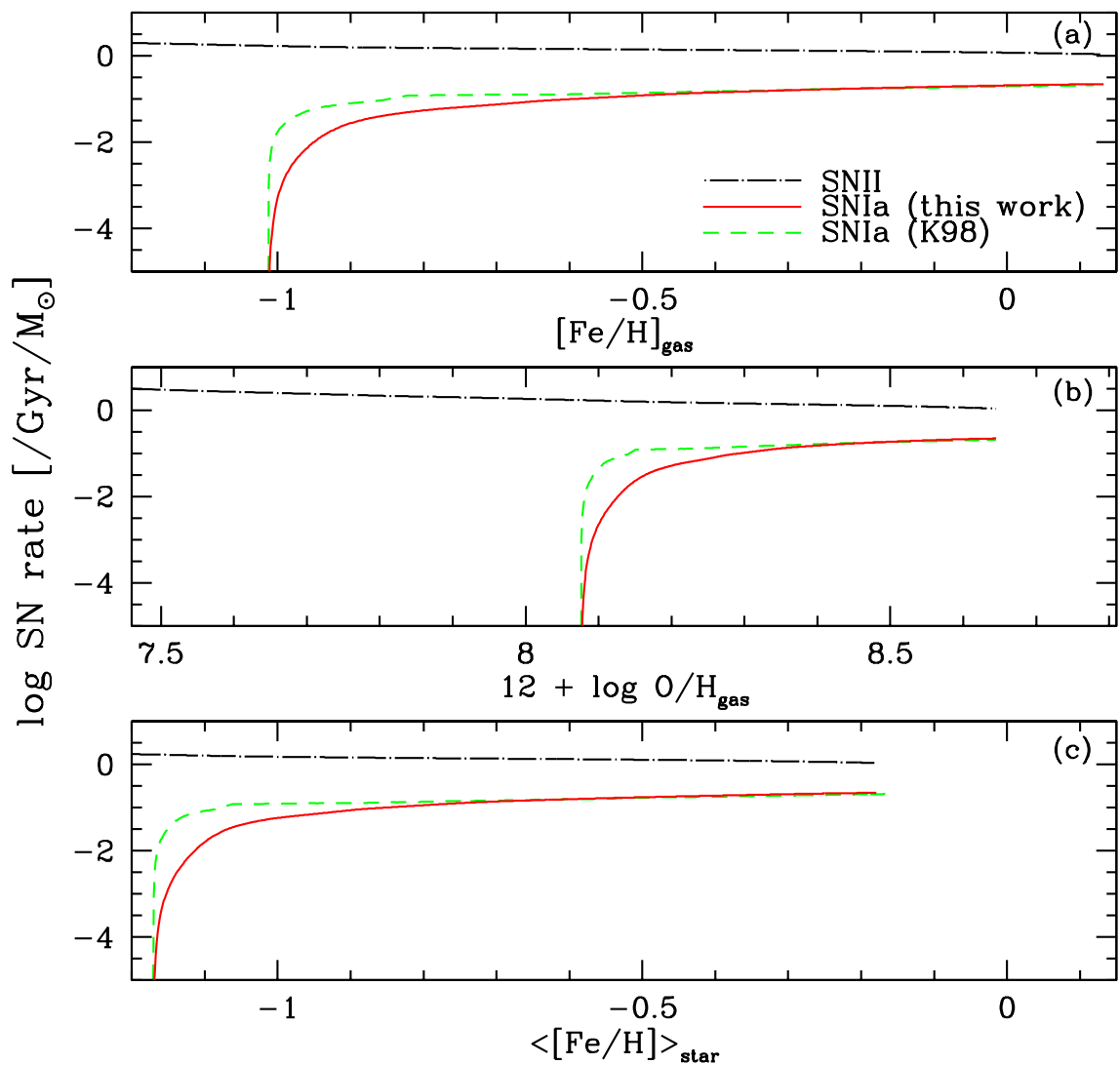}
\caption{\label{fig:zeffect}
The supernova rates against the metallicity indicators, (a) the iron abundance, (b) the gas-phase metallicity, and (c) the mean stellar metallicity, with the star formation history in the solar neighborhood.
}
\end{figure}

\begin{figure}
\center
\includegraphics[width=16cm]{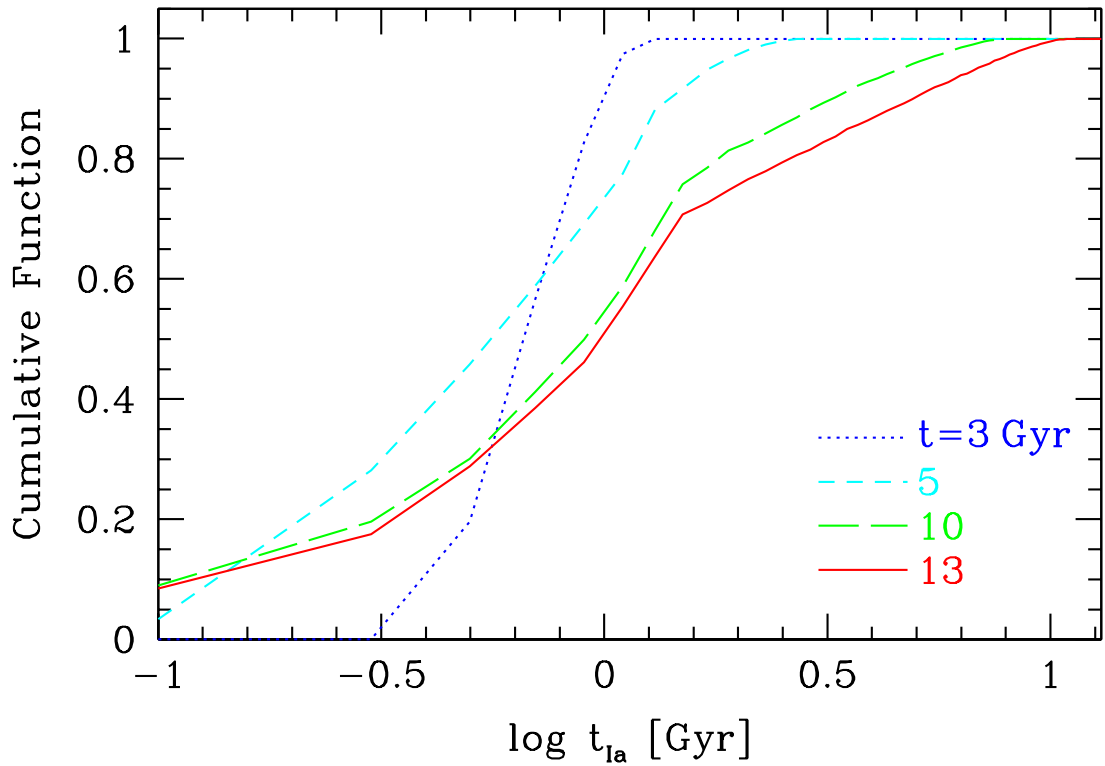}
\caption{\label{fig:snlife}
The cumulative function of SN Ia lifetimes in the solar neighborhood with our SD model at the galactic ages of $t=3$ (clue dotted line), 5 (cyan short-dashed line), 10 (green long-dashed line), and 13 Gyr (red solid line).
}
\end{figure}

\begin{figure}
\center
\includegraphics[width=16cm]{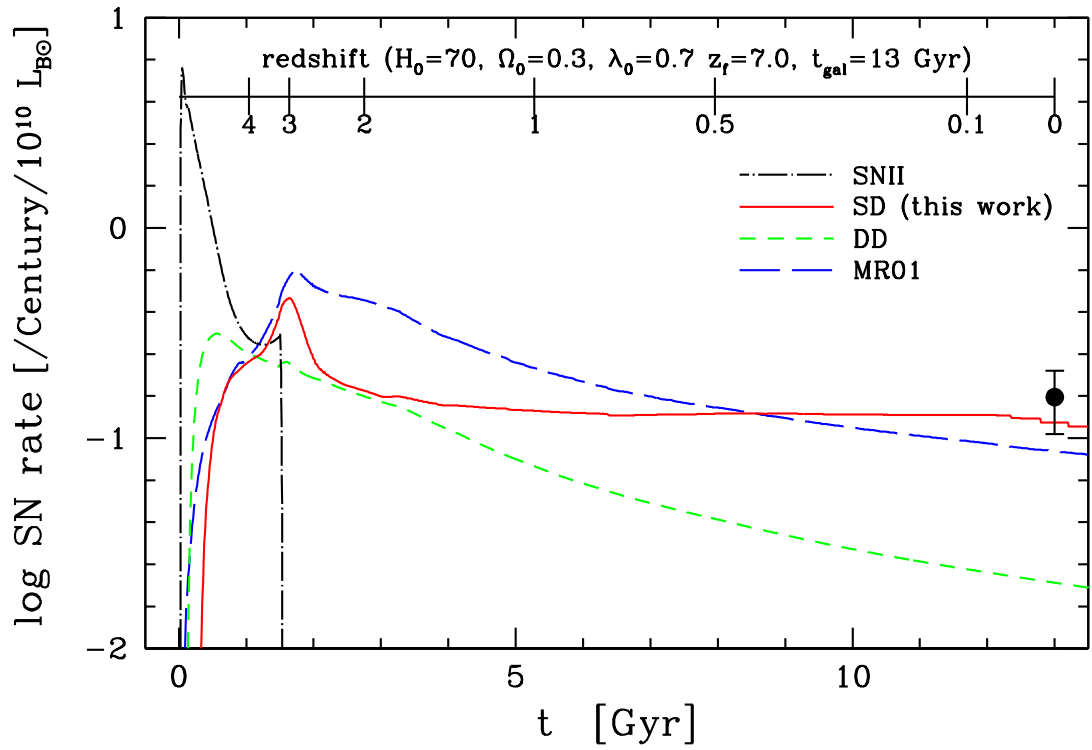}
\caption{\label{fig:ellsnr}
The SN Ia rate history in ellipticals
with different SN Ia models: 
our SD model (red solid line), DD model (green short-dashed line), and MR01 model (blue long-dashed line).
The dot-dashed line shows the SN II rate.
The dot shows the observed SN Ia rate in E-S0 type galaxies \citep{cap99}.
}
\end{figure}

\begin{figure}
\center
\includegraphics[width=15cm]{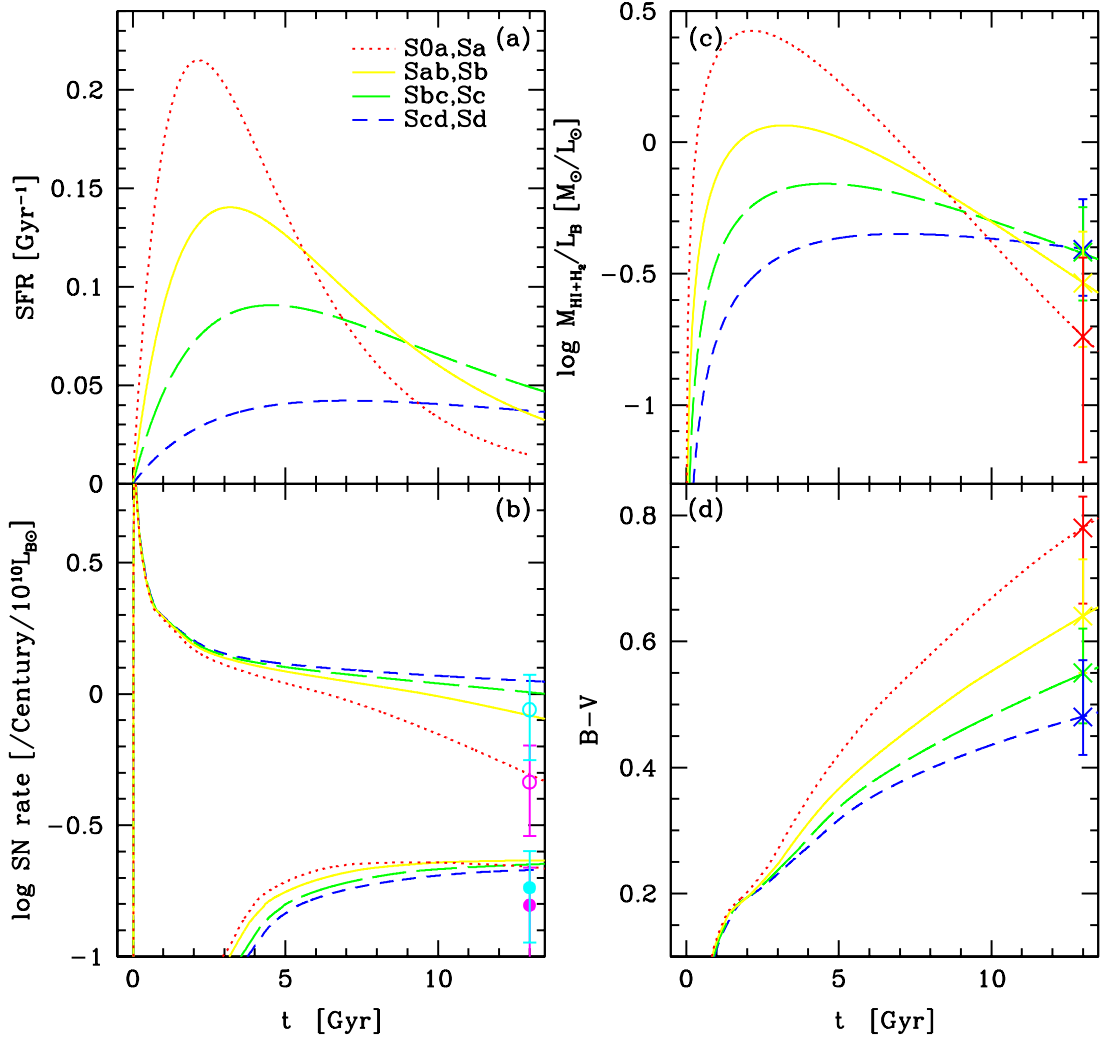}
\caption{\label{fig:spisfr}
The time evolution of (a) the SFR, (b) the SN II and Ia rates, (c) the gas fraction per luminosity, and (d) the B-V color for four types of spirals (red dotted line, S0a/Sa; yellow solid line, Sab/Sb; green long-dashed line, Sbc/Sc; blue short-dashed line, Scd/Sd).
In the panel (b), the upper and lower four lines are for the SN II and Ia rates per B-band luminosity (the supernova units, SNu), respectively.
Observational data sources are:
\cite{cap99}, open circles (SN II) and filled circles (SN Ia) of S0a-Sb and Sbc-Sd in panel (b); 
\cite{rob94}, crosses in panels (c) and (d).
}
\end{figure}

\begin{figure}
\center
\includegraphics[width=16cm]{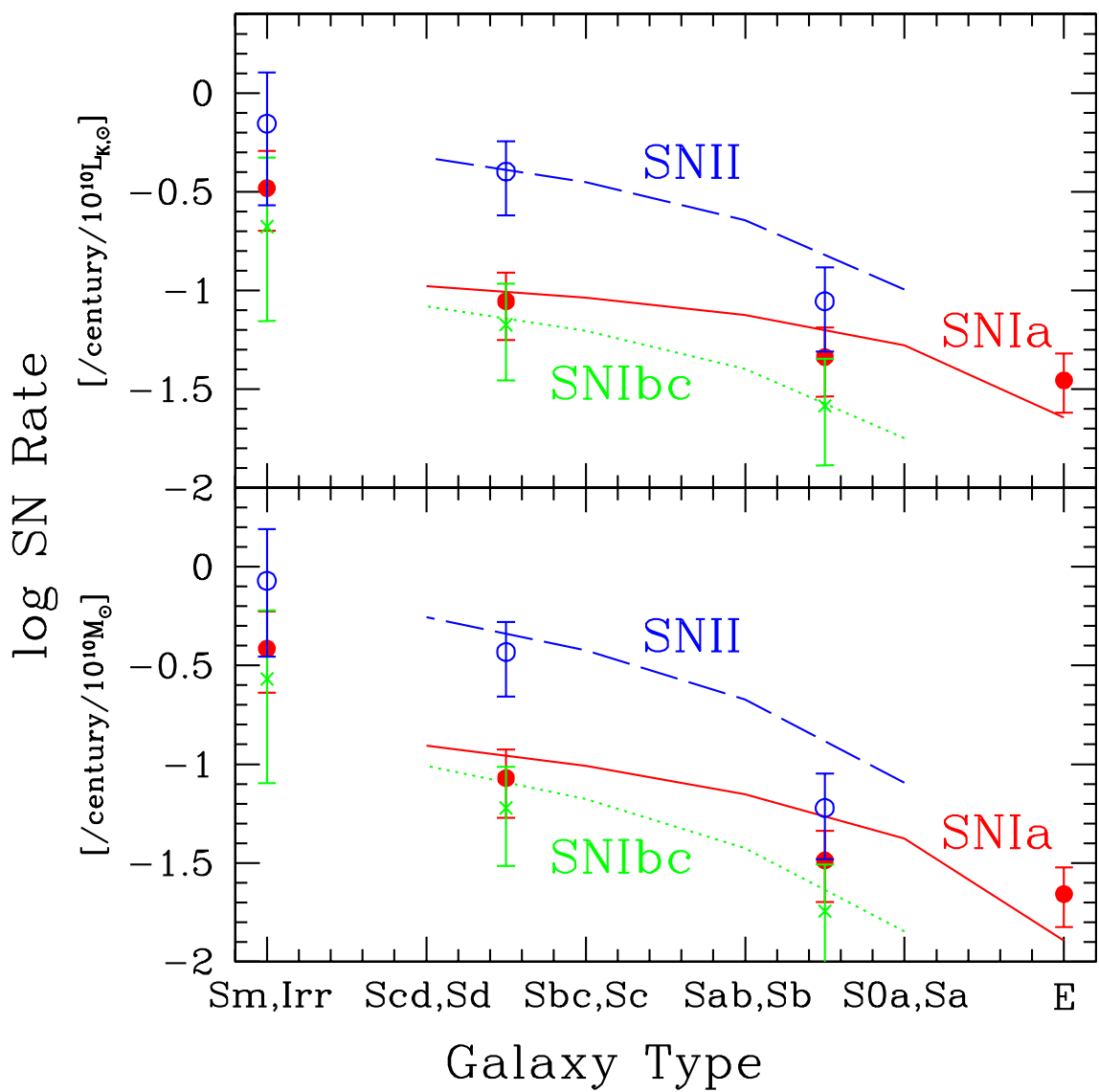}
\caption{\label{fig:snrgal}
The present supernova rates per K-band luminosity (upper panel) and per mass (lower panel) against the morphological type of galaxies for SNe II (blue dashed line), Ia (red solid line), and Ibc (green dotted line).
The observational data is taken from \cite{man05} for SNe II (open circles), Ia (filled circles), and Ibc (crosses).
}
\end{figure}

\begin{figure}
\center
\includegraphics[width=16cm]{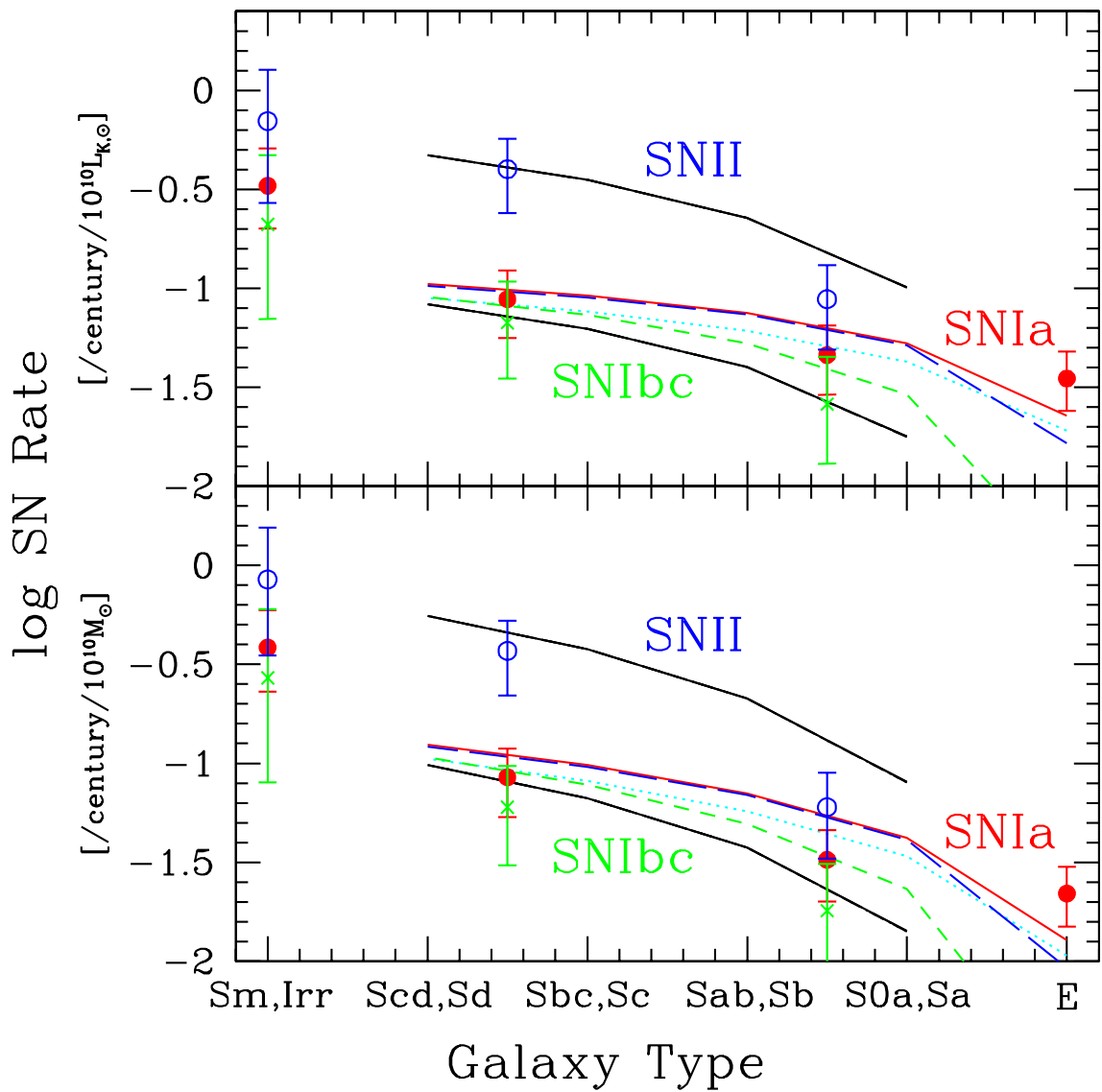}
\caption{\label{fig:snrgal-all}
Same as Figure \ref{fig:snrgal}, but for different SN Ia models: 
our SD model with (red solid line) and without (cyan dotted line) metallicity effect, DD model (green short-dashed line), and MR01 model (blue long-dashed line).
}
\end{figure}

\begin{figure}
\center
\includegraphics[width=16cm]{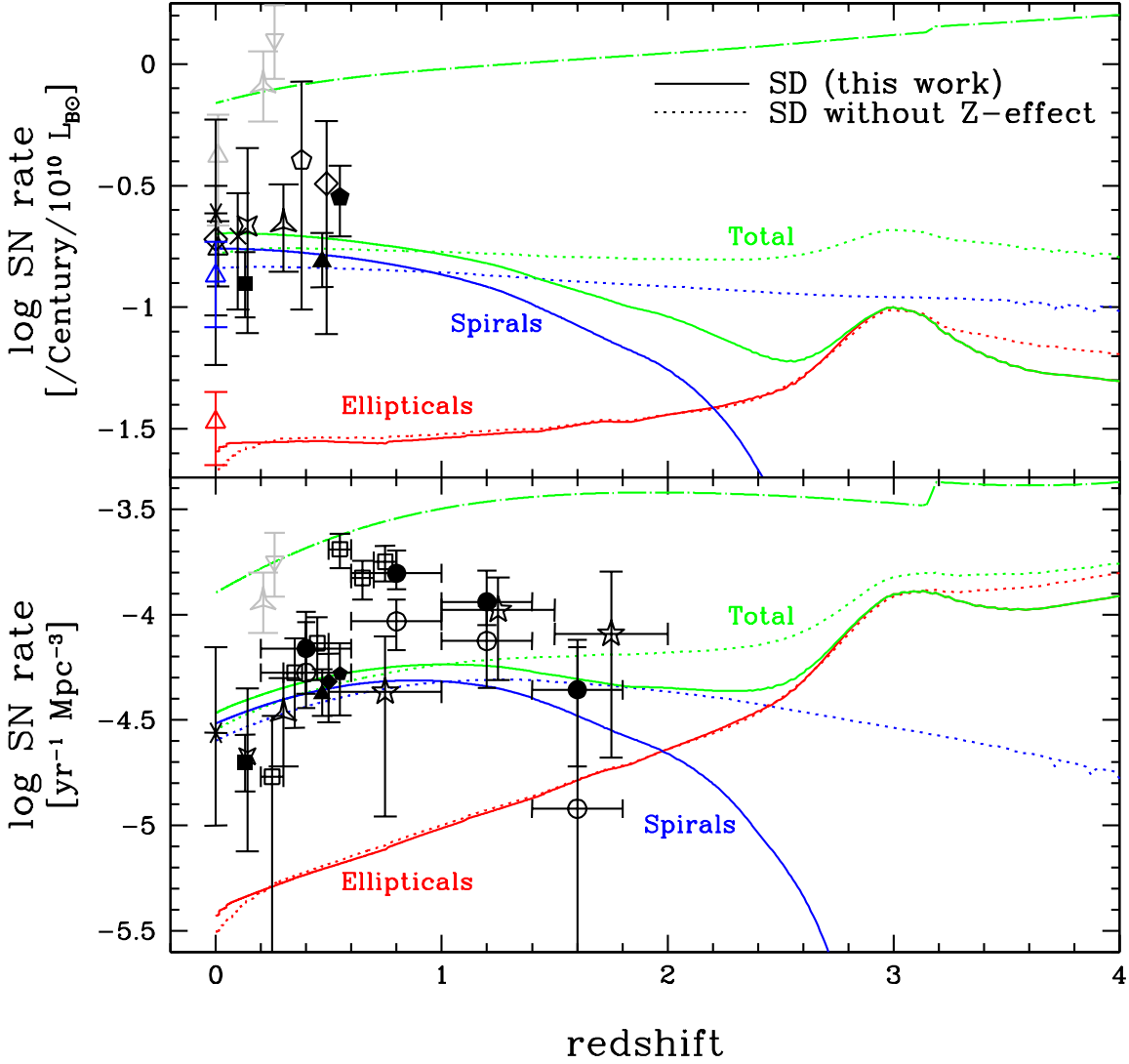}
\caption{\label{fig:snr}
The cosmic SN Ia rate histories (green lines) 
as a composite of those in spirals (blue lines) and
ellipticals (red lines)
in the supernova units (upper panel) 
and in the units of the comoving density (lower panel)
for our SD model with (solid lines) and without (dotted lines) metallicity effect.
Observational data sources are:
\citet{cap99}, open triangles (for total, Ss, and Es);
\citet{ham99}, asterisk;
\citet{har00}, four-pointed star;
\citet{pai96}, open pentagon; 
\citet{pai02}, filled pentagon;
\citet{rie96}, open diamonds;
\citet{mad03}, cross;
\citet{bla04}, filled square;
\citet{cap05}, inverted open triangle;
\citet{bot07}, three-pointed stars;
\citet{nei06}, filled triangle;
\citet{ton03}, filled diamond;
\citet{dah04}, filled circles;
\citet{bar06}, open squares;
\citet{kuz07}, open circles;
\citet{poz07}, stars.
The SN II rate history is shown with the green dot-dashed lines on the top, 
and the observed SN II rates are shown with gray color.
}
\end{figure}

\begin{figure}
\center
\includegraphics[width=16cm]{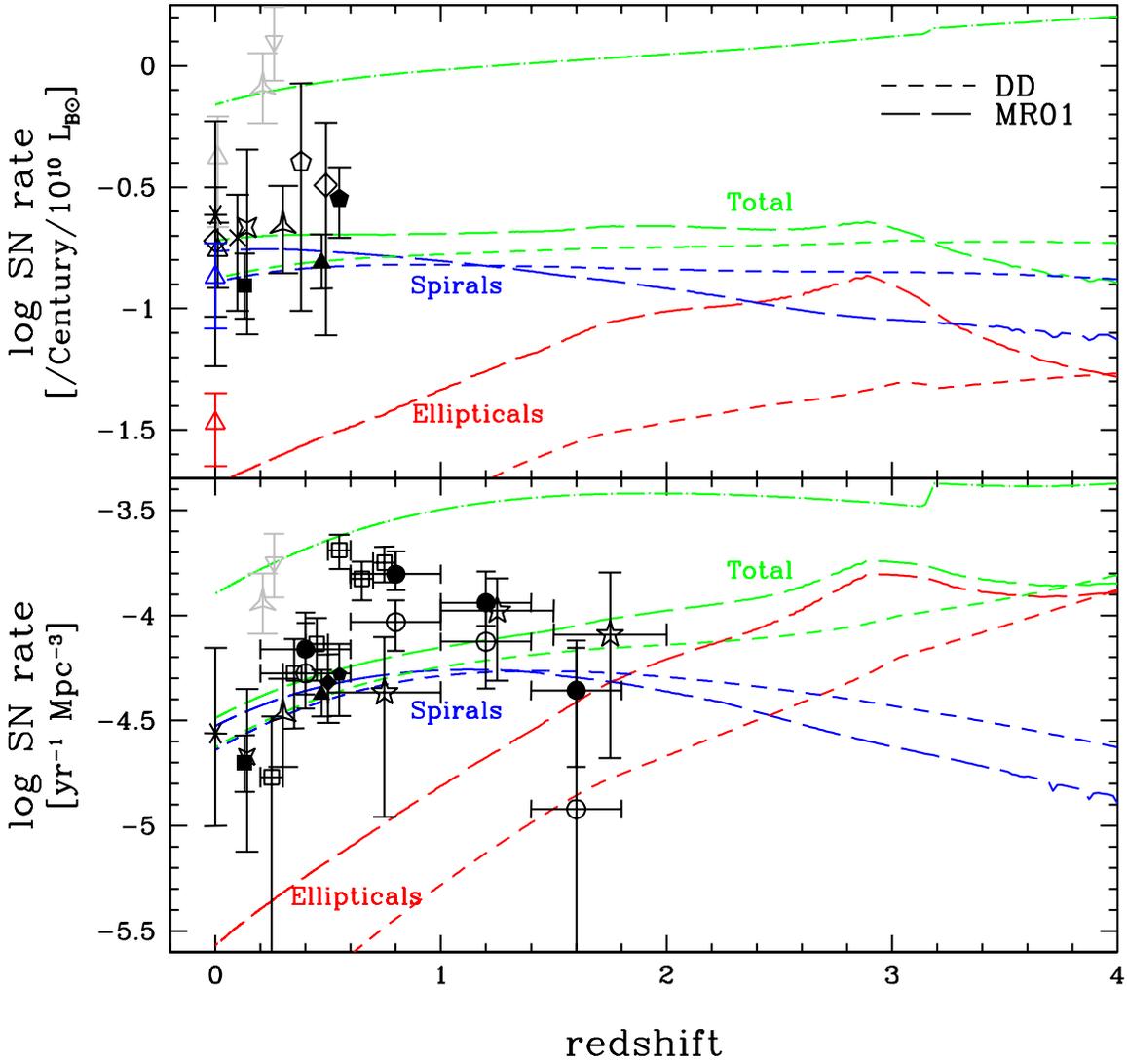}
\caption{\label{fig:snr2}
Same as Figure \ref{fig:snr}, but for DD model (short-dashed lines) and MR01 model (long-dashed lines).
}
\end{figure}

\begin{figure}
\center
\includegraphics[width=16cm]{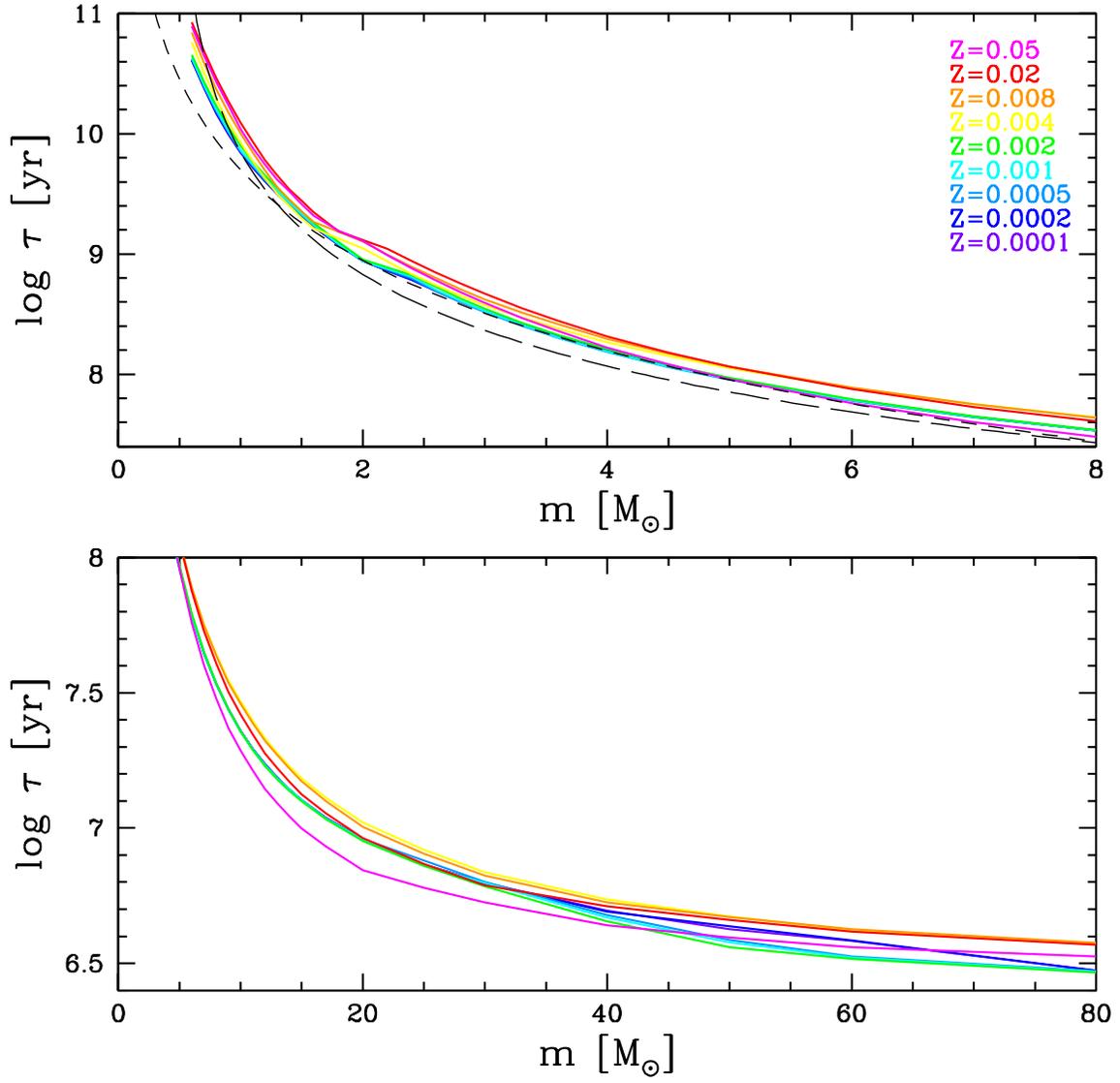}
\caption{\label{fig:lifetime}
The stellar lifetime as a function of initial mass and metallicity.
The long and short dashed-lines are for the analytic functions adopted in MR01 and \citet{hac08}, respectively.
}
\end{figure}

\begin{figure}
\center
\includegraphics[width=16cm]{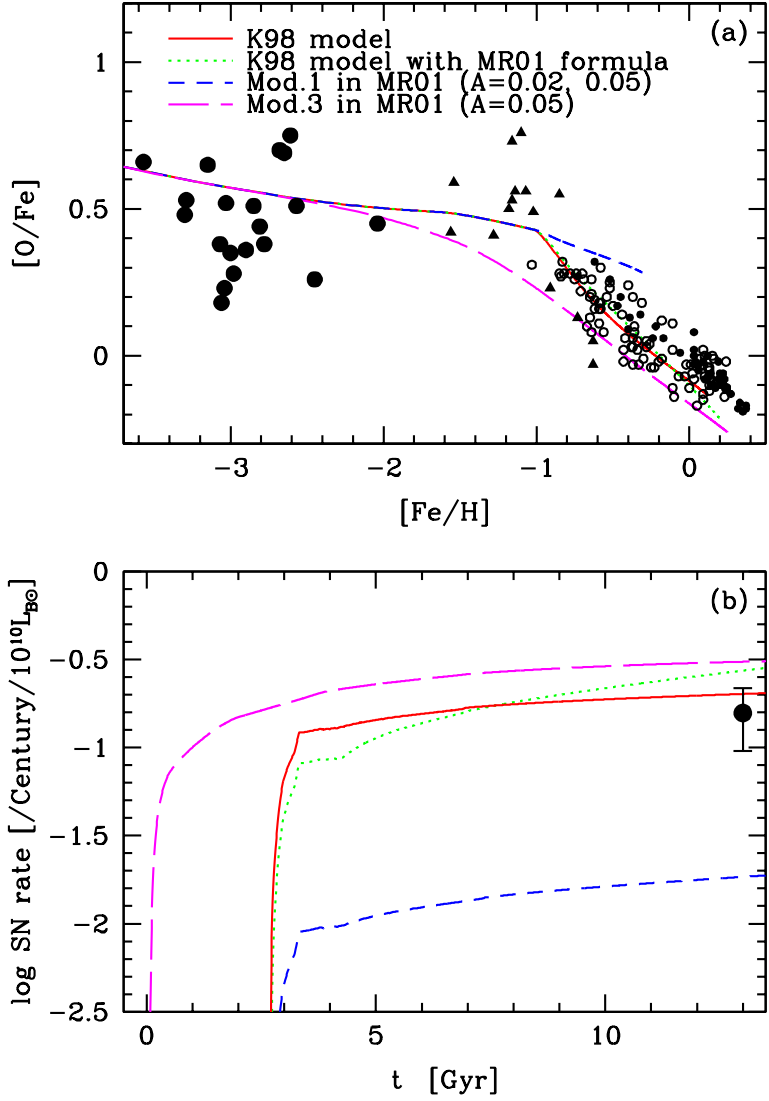}
\caption{\label{fig:ofe1}
(a) The [O/Fe]-[Fe/H] relation in the solar neighborhood for different SN Ia models.
The solid line is for our original K98 model.
The dotted line is calculated with MR01 formula Eq.[\ref{eqmat}] with the appropreate parameters for K98 model ($\gamma=-1.35, A=0.45$).
The dashed and long-dashed lines correspond to Model 1 and 3 in MR01, respectively.
Observational data sources are: 
For disk stars, \citet{edv93}, small open circles; 
thin disk stars in \citet{ben04}, small filled circles; 
accretion component in \citet{gra03}, triangles.
For halo stars, \citet{cay04}, large filled circles.
(b) The time evolution of SN Ia rate in SNu. The dot is for the observational data for S0a-Sb \citep{cap99}.
}
\end{figure}

\begin{figure}
\center
\includegraphics[width=16cm]{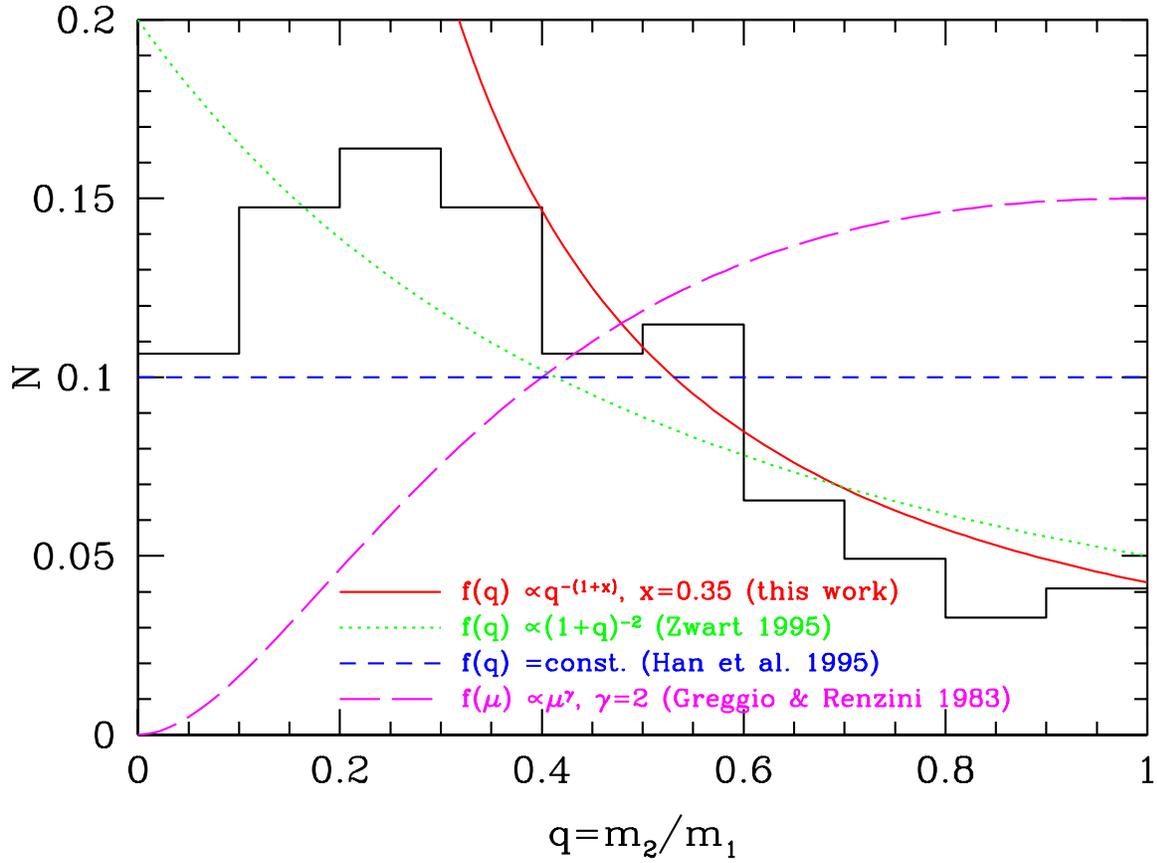}
\caption{\label{fig:binary}
The distribution function of mass ratios:
$(q)\propto m^{-(1+x)}$ with $x=0.35$ (solid line, this work),
constant $f(\mu)$ (dotted line), constant $f(q)$ (short-dashed line), and $f(\mu) \propto \mu^\gamma$ with $\gamma=2$ (long-dashed line, MR01).
The histogram shows the observational data \citep{duq91}.
}
\end{figure}

\begin{figure}
\center
\includegraphics[width=16cm]{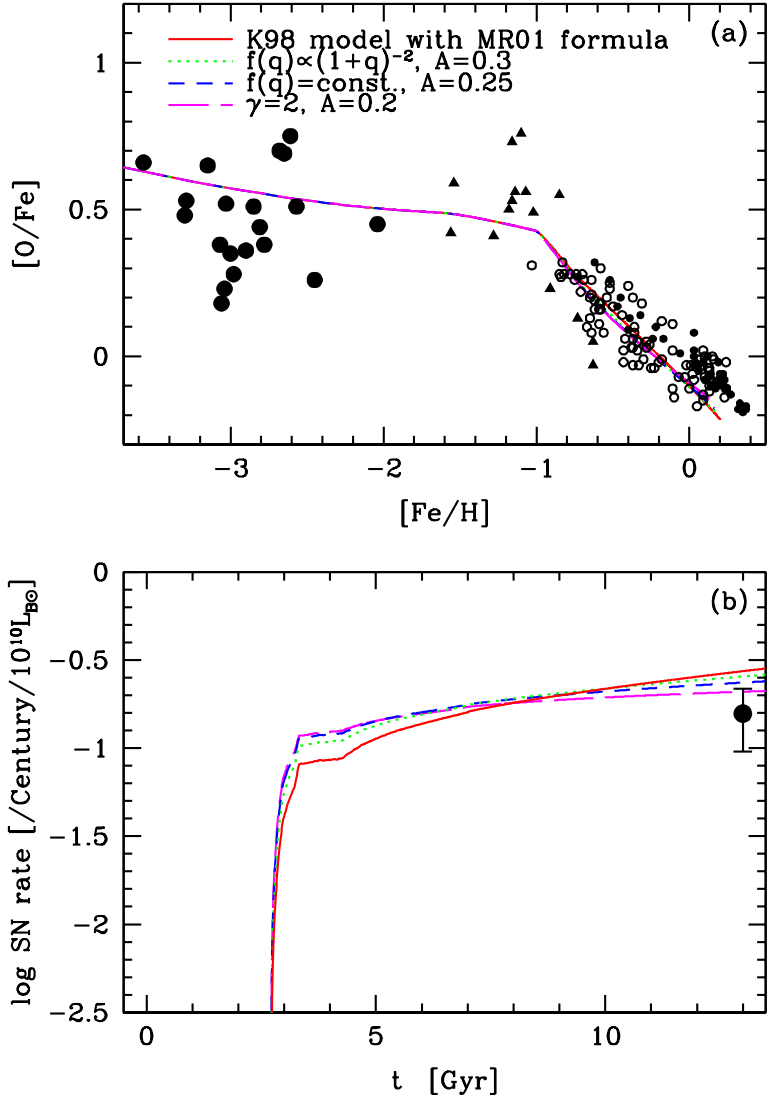}
\caption{\label{fig:ofe2}
The same as Figure \ref{fig:ofe1} but for different SN Ia models.
The solid line is calculated with MR01 formula Eq.[\ref{eqmat}] with the appropreate parameters for K98 model.
The dotted and short-dashed lines are for the different functions of mass ratios, constant $f(\mu)$ and constant $f(q)$, respectively.
The long-dashed line is for the same $f(\mu)$ as in MR01, but with our parameters of binary companions.
}
\end{figure}

\noindent
(4) We also provide the cosmic supernova rate history as a composite of those in spirals and ellipticals.
The cosmic SN Ia rate in comoving density shows a peak at $z\sim1$ and decreases toward higher redshift in our model.
In contrast, this rate increases toward higher redshifts in the DD and MR01-like models.
Because of the metallicity effect, i.e., because of the lack of winds from WDs in the binary systems at [Fe/H] $< -1.1$, the SN Ia rate in the low-metallicity systems, e.g., high-z spiral galaxies, is supposed to be very small in our model.
This metallicity effect will appear only around [Fe/H] $\sim -1$ (Fig. \ref{fig:zeffect}).
In contrast, the SN Ia rate in such systems is as high as present in the DD and MR01-like models.
In our models, at $z \gtsim 1$, SNe Ia will be observed only in the systems that have evolved with a short timescale of chemical enrichment. This suggests that the evolution effect in the supernova cosmology can be small.

\acknowledgments

This work has been supported in part by World Premier International
Research Center Initiative (WPI Initiative), the Ministry of Education, Culture, Sports, Science, and Technology (MEXT) of Japan, and by the
Grant-in-Aid for Scientific Research of the Japan Society for Promotion of Science (JSPS) (18104003, 18540231, 20244035, 20540226) and MEXT (19047004, 20040004),
and also by the National Science Foundation under Grant No. PHY05-51164 at Kavli Institute for Theoretical Physics of USA.
C.K. thank to the JSPS for a financial support. 
We would like to thank
I. Hachisu, M. Kato, M. Tanaka, F. Matteucci, S. Recchi, M. Della Valle, P. Podsiadlowski, B. Schmidt, R. Ellis, L. Greggio, and A. Renzini 
for fruitful discussion.
We are particularly grateful to the late Prof. B. E. J. Pagel for generous suggestion and encouragement.

\newpage

\newpage

\newpage

\newpage
\begin{deluxetable}{lrrrrrrrr|rr|r|r}
\tablenum{1}
\tablecaption{\label{tab:snia}
The SN Ia rate of simple stellar population in [$10^{-4}~M_\odot^{-1}$] as a function of companion's mass $m_{\rm d}$ [$M_\odot$].}
\tablewidth{0pt}
\footnotesize
\tablehead{
& \multicolumn{8}{c}{this work} & \multicolumn{2}{|c|}{K98} & MR01 & DD \\
& \multicolumn{2}{c}{$Z=0.002$} & \multicolumn{2}{c}{$Z=0.004$} & \multicolumn{2}{c}{$Z=0.02$} & \multicolumn{2}{c|}{$Z=0.05$} & \multicolumn{2}{c|}{[Fe/H] $\ge -1.1$} & & \\
$m_{\rm d}$ &
${\cal R}_{\rm Ia,m}$ & ${\cal R}_{\rm Ia,m}$ & 
${\cal R}_{\rm Ia,m}$ & ${\cal R}_{\rm Ia,m}$ & 
${\cal R}_{\rm Ia,m}$ & ${\cal R}_{\rm Ia,m}$ & 
${\cal R}_{\rm Ia,m}$ & ${\cal R}_{\rm Ia,m}$ & 
${\cal R}_{\rm Ia,m}$ & ${\cal R}_{\rm Ia,m}$ & 
${\cal R}_{\rm Ia,m}$ & ${\cal R}_{\rm Ia,m}$ \\
& RG & MS & RG & MS & RG & MS & RG & MS & RG & MS & &  
}
\startdata
   0.6&      0&      0&      0&      0&      0&      0&      0&      0&      0&      0&  1.439&      0\\
   0.7&      0&      0&      0&      0&      0&      0&      0&      0&      0&      0&  1.976&      0\\
   0.8&      0&      0&      0&      0&      0&      0&  2.874&      0&      0&      0&  2.550&      0\\
   0.9&  2.193&      0&  2.193&      0&  2.193&      0&  6.878&      0&  2.841&      0&  3.283&  0.479\\
   1.0&  5.966&      0&  5.966&      0&  5.966&      0&  5.966&      0&  7.729&      0&  4.020&  0.864\\
   1.1&  5.246&      0&  5.246&      0&  5.246&      0&  5.246&      0&  6.796&      0&  4.922&  1.244\\
   1.2&  4.664&      0&  4.664&      0&  4.664&      0&  4.664&      0&  6.043&      0&  5.787&  1.770\\
   1.3&  4.187&      0&  4.187&      0&  4.187&      0&  4.187&      0&  5.424&      0&  6.862&  2.148\\
   1.4&  3.788&      0&  3.788&      0&  3.788&      0&  3.788&      0&  4.907&      0&  7.847&  2.598\\
   1.5&  2.802&      0&  3.451&      0&  3.451&      0&  3.451&      0&  3.630&      0&  8.977&  2.755\\
   1.6&      0&      0&  3.163&      0&  3.163&      0&  3.163&      0&      0&      0&  7.822&  2.927\\
   1.7&      0&      0&  2.915&      0&  2.915&      0&  2.915&      0&      0&      0&  6.751&  2.886\\
   1.8&      0&  0.218&  2.698&  0.218&  2.698&  0.218&  2.698&  0.218&      0&  0.756&  5.870&  2.347\\
   1.9&      0&  3.247&  2.508&  3.247&  2.508&  3.247&  2.508&  3.247&      0&  11.27&  5.140&  1.460\\
   2.0&      0&  3.030&  2.156&  3.030&  2.340&  3.030&  2.340&  3.030&      0&  10.52&  4.527&  1.512\\
   2.1&      0&  2.837&      0&  2.837&  2.191&  2.837&  2.191&  2.837&      0&  9.845&  4.009&  1.586\\
   2.2&      0&  2.664&      0&  2.664&  2.058&  2.664&  2.058&  2.664&      0&  9.246&  3.567&  1.947\\
   2.3&      0&  2.509&      0&  2.509&  1.938&  2.509&  1.938&  2.509&      0&  8.708&  3.187&  2.458\\
   2.4&      0&  2.369&      0&  2.369&  1.830&  2.369&  1.830&  2.369&      0&  8.221&  2.859&  2.529\\
   2.5&      0&  2.242&      0&  2.242&  1.732&  2.242&  1.732&  2.242&      0&  7.780&  2.574&  2.616\\
   2.6&      0&  2.036&      0&  2.126&  1.642&  2.126&  1.642&  2.126&      0&  7.067&  2.324&  2.695\\
   2.7&      0&      0&      0&  2.020&  1.561&  2.020&  1.561&  2.020&      0&      0&  2.105&  2.789\\
   2.8&      0&      0&      0&  1.924&  1.486&  1.924&  1.486&  1.924&      0&      0&  1.912&  2.833\\
   2.9&      0&      0&      0&  1.835&  1.417&  1.835&  1.417&  1.835&      0&      0&  1.740&  2.875\\
   3.0&      0&      0&      0&  1.753&  0.095&  1.753&  1.354&  1.753&      0&      0&  1.587&  2.587\\
   3.3&      0&      0&      0&  1.541&      0&  1.541&  1.190&  1.541&      0&      0&  1.219&  3.090\\
   3.6&      0&      0&      0&  1.370&      0&  1.370&      0&  1.370&      0&      0&  0.948&  3.049\\
   4.0&      0&      0&      0&  0.216&      0&  1.189&      0&  1.189&      0&      0&  0.688&  3.407\\
   4.3&      0&      0&      0&      0&      0&  1.078&      0&  1.078&      0&      0&  0.545&  2.615\\
   4.6&      0&      0&      0&      0&      0&  0.984&      0&  0.984&      0&      0&  0.433&  2.162\\
   5.0&      0&      0&      0&      0&      0&  0.879&      0&  0.879&      0&      0&  0.319&  0.733\\
   5.5&      0&      0&      0&      0&      0&  0.734&      0&  0.773&      0&      0&  0.216&  0.041\\
   6.0&      0&      0&      0&      0&      0&      0&      0&  0.228&      0&      0&  0.142&  0.003\\
   6.5&      0&      0&      0&      0&      0&      0&      0&      0&      0&      0&  0.088&      0\\
   7.0&      0&      0&      0&      0&      0&      0&      0&      0&      0&      0&  0.049&      0\\
   7.5&      0&      0&      0&      0&      0&      0&      0&      0&      0&      0&  0.020&      0\\
   8.0&      0&      0&      0&      0&      0&      0&      0&      0&      0&      0&      0&      0\\
\enddata
\end{deluxetable}

\begin{deluxetable}{rrrrrrrrrrrr}
\tablenum{2}
\tablecaption{\label{tab:snia2}
The SN Ia rate of simple stellar population in [$10^{-4}~M_\odot^{-1}$] as a function of time $t_{\rm Ia}$ [Gyr].}
\tablewidth{0pt}
\footnotesize
\tablehead{
\multicolumn{12}{c}{this work} \\
\multicolumn{3}{c}{$Z=0.002$} & \multicolumn{3}{c}{$Z=0.004$} & \multicolumn{3}{c}{$Z=0.02$} & \multicolumn{3}{c}{$Z=0.05$} \\
$t_{\rm Ia}$ & ${\cal R}_{\rm Ia,t}$ & ${\cal R}_{\rm Ia,t}$ &
$t_{\rm Ia}$ & ${\cal R}_{\rm Ia,t}$ & ${\cal R}_{\rm Ia,t}$ &
$t_{\rm Ia}$ & ${\cal R}_{\rm Ia,t}$ & ${\cal R}_{\rm Ia,t}$ &
$t_{\rm Ia}$ & ${\cal R}_{\rm Ia,t}$ & ${\cal R}_{\rm Ia,t}$ \\
& RG & MS & & RG & MS & & RG & MS & & RG & MS
}
\startdata
 45.15&      0&      0&  57.50&      0&      0&  84.76&      0&      0&  78.29&      0&      0\\
 27.16&      0&      0&  32.39&      0&      0&  48.76&      0&      0&  44.78&      0&      0\\
 17.08&      0&      0&  19.56&      0&      0&  29.17&      0&      0&  27.04&  0.021&      0\\
 11.33&  0.048&      0&  12.53&  0.039&      0&  18.59&  0.026&      0&  17.24&  0.086&      0\\
 7.974&  0.219&      0&  8.413&  0.181&      0&  12.36&  0.119&      0&  11.00&  0.124&      0\\
 5.891&  0.305&      0&  5.929&  0.253&      0&  8.575&  0.166&      0&  7.596&  0.193&      0\\
 4.533&  0.401&      0&  4.265&  0.338&      0&  6.026&  0.229&      0&  5.563&  0.271&      0\\
 3.563&  0.487&      0&  3.165&  0.480&      0&  4.501&  0.327&      0&  4.151&  0.374&      0\\
 2.812&  0.587&      0&  2.519&  0.649&      0&  3.462&  0.441&      0&  3.322&  0.491&      0\\
 2.271&  0.558&      0&  1.998&  0.776&      0&  2.783&  0.562&      0&  2.607&  0.557&      0\\
 1.807&      0&      0&  1.629&  1.217&      0&  2.234&  0.675&      0&  2.082&  0.777&      0\\
 1.502&      0&      0&  1.478&  2.083&      0&  1.845&  0.840&      0&  1.793&  1.109&      0\\
 1.262&      0&  0.099&  1.349&  2.127&  0.172&  1.540&  1.253&  0.101&  1.557&  1.418&  0.115\\
 1.063&      0&  1.812&  1.225&  2.164&  2.801&  1.415&  2.129&  2.756&  1.412&  1.864&  2.413\\
 0.903&      0&  2.678&  1.117&  1.758&  2.470&  1.305&  2.166&  2.803&  1.287&  1.597&  2.068\\
 0.836&      0&  4.513&  0.979&      0&  2.238&  1.198&  2.196&  2.842&  1.119&  1.421&  1.840\\
 0.777&      0&  4.581&  0.864&      0&  2.498&  1.105&  1.896&  2.455&  0.979&  1.599&  2.070\\
 0.720&      0&  4.637&  0.766&      0&  2.774&  0.981&  1.691&  2.189&  0.862&  1.790&  2.317\\
 0.669&      0&  3.804&  0.683&      0&  3.066&  0.876&  1.840&  2.382&  0.763&  1.973&  2.554\\
 0.596&      0&  3.279&  0.612&      0&  3.375&  0.783&  1.993&  2.580&  0.676&  2.164&  2.802\\
 0.533&      0&  3.374&  0.550&      0&  3.701&  0.702&  2.160&  2.796&  0.603&  2.372&  3.070\\
 0.475&      0&      0&  0.497&      0&  4.045&  0.630&  2.330&  3.017&  0.538&  2.587&  3.349\\
 0.425&      0&      0&  0.450&      0&  4.408&  0.568&  2.554&  3.306&  0.482&  2.852&  3.692\\
 0.386&      0&      0&  0.409&      0&  4.789&  0.514&  2.789&  3.610&  0.434&  3.132&  4.055\\
 0.351&      0&      0&  0.374&      0&  6.163&  0.467&  0.239&  4.431&  0.392&  3.912&  5.063\\
 0.268&      0&      0&  0.296&      0&  6.864&  0.356&      0&  4.899&  0.295&  4.371&  5.658\\
 0.212&      0&      0&  0.239&      0&  8.623&  0.278&      0&  6.398&  0.228&      0&  7.479\\
 0.161&      0&      0&  0.184&      0&  1.847&  0.206&      0&  7.788&  0.167&      0&  9.124\\
 0.135&      0&      0&  0.157&      0&      0&  0.171&      0&  10.41&  0.137&      0&  12.17\\
 0.114&      0&      0&  0.135&      0&      0&  0.144&      0&  12.54&  0.114&      0&  14.86\\
 0.094&      0&      0&  0.112&      0&      0&  0.116&      0&  15.43&  0.090&      0&  18.65\\
 0.075&      0&      0&  0.092&      0&      0&  0.093&      0&  17.97&  0.071&      0&  23.34\\
 0.062&      0&      0&  0.077&      0&      0&  0.075&      0&      0&  0.057&      0&  9.631\\
 0.052&      0&      0&  0.065&      0&      0&  0.063&      0&      0&  0.048&      0&      0\\
 0.045&      0&      0&  0.056&      0&      0&  0.053&      0&      0&  0.040&      0&      0\\
 0.039&      0&      0&  0.050&      0&      0&  0.046&      0&      0&  0.035&      0&      0\\
 0.034&      0&      0&  0.044&      0&      0&  0.041&      0&      0&  0.030&      0&      0\\
\enddata
\end{deluxetable}

\begin{deluxetable}{rrrrrr|r|r}
\tablenum{3}
\tablecaption{\label{tab:snia3}
The SN Ia rate of simple stellar population in [$10^{-4}~M_\odot^{-1}$] as a function of time $t_{\rm Ia}$ [Gyr].}
\tablewidth{0pt}
\footnotesize
\tablehead{
\multicolumn{6}{c|}{K98} & MR01 & DD \\
\multicolumn{3}{c}{$Z=0.002$} & \multicolumn{3}{c|}{$Z=0.02$} & & \\
$t_{\rm Ia}$ & ${\cal R}_{\rm Ia,t}$ & ${\cal R}_{\rm Ia,t}$ &
$t_{\rm Ia}$ & ${\cal R}_{\rm Ia,t}$ & ${\cal R}_{\rm Ia,t}$ &
${\cal R}_{\rm Ia,t}$ & ${\cal R}_{\rm Ia,t}$ \\
& RG & MS & & RG & MS & & 
}
\startdata
 45.15&      0&      0&  84.76&      0&      0&  0.003&      0\\
 27.16&      0&      0&  48.76&      0&      0&  0.007&      0\\
 17.08&      0&      0&  29.17&      0&      0&  0.017&      0\\
 11.33&  0.062&      0&  18.59&  0.034&      0&  0.039&  0.006\\
 7.974&  0.284&      0&  12.36&  0.154&      0&  0.080&  0.017\\
 5.891&  0.395&      0&  8.575&  0.215&      0&  0.156&  0.039\\
 4.533&  0.519&      0&  6.026&  0.297&      0&  0.284&  0.087\\
 3.563&  0.630&      0&  4.501&  0.423&      0&  0.535&  0.168\\
 2.812&  0.760&      0&  3.462&  0.571&      0&  0.913&  0.302\\
 2.271&  0.722&      0&  2.783&  0.592&      0&  1.463&  0.449\\
 1.807&      0&      0&  2.234&      0&      0&  1.668&  0.624\\
 1.502&      0&      0&  1.845&      0&      0&  1.945&  0.832\\
 1.262&      0&  0.345&  1.540&      0&  0.351&  2.727&  1.090\\
 1.063&      0&  6.289&  1.415&      0&  9.565&  4.363&  1.239\\
 0.903&      0&  9.295&  1.305&      0&  9.730&  4.189&  1.399\\
 0.836&      0&  15.66&  1.198&      0&  9.865&  4.017&  1.589\\
 0.777&      0&  15.90&  1.105&      0&  8.520&  3.287&  1.794\\
 0.720&      0&  16.09&  0.981&      0&  7.596&  2.780&  2.144\\
 0.669&      0&  13.20&  0.876&      0&  8.267&  2.875&  2.543\\
 0.596&      0&  11.38&  0.783&      0&  8.956&  2.963&  3.012\\
 0.533&      0&  11.71&  0.702&      0&  9.293&  3.057&  3.543\\
 0.475&      0&      0&  0.630&      0&      0&  3.143&  4.165\\
 0.425&      0&      0&  0.568&      0&      0&  3.285&  4.868\\
 0.386&      0&      0&  0.514&      0&      0&  3.425&  5.657\\
 0.351&      0&      0&  0.467&      0&      0&  4.014&  6.542\\
 0.268&      0&      0&  0.356&      0&      0&  3.873&  9.824\\
 0.212&      0&      0&  0.278&      0&      0&  4.426&  14.24\\
 0.161&      0&      0&  0.206&      0&      0&  4.510&  22.33\\
 0.135&      0&      0&  0.171&      0&      0&  5.267&  25.27\\
 0.114&      0&      0&  0.144&      0&      0&  5.522&  27.56\\
 0.094&      0&      0&  0.116&      0&      0&  5.598&  12.86\\
 0.075&      0&      0&  0.093&      0&      0&  5.274&  1.004\\
 0.062&      0&      0&  0.075&      0&      0&  4.807&  0.086\\
 0.052&      0&      0&  0.063&      0&      0&  4.066&      0\\
 0.045&      0&      0&  0.053&      0&      0&  2.949&      0\\
 0.039&      0&      0&  0.046&      0&      0&  1.532&      0\\
 0.034&      0&      0&  0.041&      0&      0&      0&      0\\
\enddata
\end{deluxetable}


\begin{thebibliography}{}
\setlength{\parskip}{-2pt}

\bibitem[Aldering et al.(2006)]{ald06}
Aldering, G. et al. 2006, \apj, 650, 510

\bibitem[Aoki et al.(2009)]{aok09}
Aoki, W., et al. 2009, \aap, 502, 569

\bibitem[Arnett(1996)]{arnett1996}
Arnett, W. D. 1996, Supernovae and Nucleosynthesis (Princeton Univ. Press)

\bibitem[Badenes et al.(2007)]{bad07}
Badenes, C., Hughes, J. P., Bravo, E., \& Langer, N. 2007, \apj, 662, 472

\bibitem[Badenes et al.(2009)]{bad09}
Badenes, C., Harris, J., Zaritsky, D. \&Prieto, J. L. 2009, \apj, 700, 727

\bibitem[Barris \& Tonry(2006)]{bar06}
Barris, B. J. \& Tonry, J. L. 2006, \apj, 637, 427

\bibitem[Belczynski et al.(2005)]{bel05}
Belczynski, K., Bulik, T., \& Rutter A. J. 2005, \apj, 629, 915

\bibitem[Bensby et al.(2003)]{ben03}
Bensby, T., Feltzing, S., Lundstr\"{o}m, I. 2003, \aap, 410, 527

\bibitem[Bensby et al.(2004)]{ben04}
Bensby, T., Feltzing, S., Lundstr\"{o}m, I. 2004, \aap, 415, 155

\bibitem[Binggeli et al.(1984)]{bin84} 
Binggeli, B., Sandage, A., \& Tarenghi, M. 1984, \aj, 89, 64

\bibitem[Blanc et al.(2004)]{bla04}
Blanc, G. et al. 2004, \aap, 423, 881

\bibitem[Boesgaard et al.(1999)]{boe99}
Boesgaard, A. M., King, J. R., Deliyannis, C. P., Vogt, S. S. 1999, \aj, 117, 492

\bibitem[Botticella et al.(2008)]{bot07}
Botticella, M. T. et al. 2008, \aap, 479, 49

\bibitem[Calura \& Matteucci(2006)]{cal06}
Calura, F. \& Matteucci, F. 2006, \apj, 652, 889

\bibitem[Cappellaro et al.(1999)]{cap99}
Cappellaro, E., Evans, R., \& Turatto, M. 1997, \aap, 351, 459

\bibitem[Cappellaro et al.(2005)]{cap05}
Cappellaro, E., et al. 2005, \aap, 430, 83

\bibitem[Cayrel et al.(2004)]{cay04}
Cayrel, R. et al. 2004, \aap, 416, 1117

\bibitem[Cooper, Newman \& Yan(2009)]{coo09}
Cooper, M. C., Newman, J. A., \& Yan R. 2009, astro-ph/0901.4338

\bibitem[Dahlen et al.(2004)]{dah04}
Dahlen, T. et al. 2004, \apj, 613, 189

\bibitem[Della Valle et al.(2005)]{del05}
Della Valle, M., Panagia, N., Padovani, P., Cappellaro, E., Mannucci, F., \& Turatto, M. 2005, \apj, 629, 750

\bibitem[Deng et al.(2004)]{den04}
Deng, J. et al. 2004, \apj, 605, L37

\bibitem[Dessauges-Zavadsky et al.(2007)]{des07}
Dessauges-Zavadsky, M., Calura, F., Prochaska, J. X., D'Odorico, S., \& Matteucci, F. 2007, \aap, 470, 431

\bibitem[Duquennoy \& Mayor(1991)]{duq91}
Duquennoy, A., \& Mayor, M. 1991, \aap, 248, 485

\bibitem[Edvardsson et al.(1993)]{edv93} 
Edvardsson, B., Andersen, J., Gustafsson, B., Lambert, D. L., Nissen, P. E.,
 \& Tomkin, J. 1993, \aap, 275, 101

\bibitem[Feltzing, Fohlman, \& Bensby(2007)]{fel07}
Feltzing, S., Fohlman, M., \& Bensby, T. 2007, \aap, 467, 665

\bibitem[Filippenko(1997)]{alex1997} 
Filippenko, A.V. 1997, \araa, 35, 309

\bibitem[F\"oerster \& Schawinski (2008)]{foe08}
F\"orster, F. \& Schawinski, K. 2008, \mnras, 388, L74

\bibitem[Frebel et al.(2009)]{fre09}
Frebel, A., Simon, J. D., Geha, M., \& Willman, B., 2009, astro-ph/0902.2395

\bibitem[Frebel et al.(2009b)]{fre09b}
Frebel, A. et al. 2009, talk at IAU Symposium 265

\bibitem[Gallagher et al.(2005)]{gal05}
Gallagher, J. S., Garnavich, P. M., Berlind, P., Challis, P, Jha, S., \& Kirshner, R. P. 2005, \apj, 634, 210

\bibitem[Gallagher et al.(2008)]{gal08}
Gallagher, J. S., Garnavich, P. M., Caldwell, N., Kirshner, R. P., Jha, S. W., Li W., Ganeshalingam, M., \& Filippenko, A. V. 2008, \apj, 685, 752

\bibitem[Geier et al.(2007)]{gei07}
Geier, S., Nesslinger, S., Heber, U., Przybilla, N., Napiwotzki, R., Kudritzki, R.-P. 2007, \aap, 464, 299

\bibitem[Gratton(1989)]{gra89}
Gratton 1989, \aap, 208, 171

\bibitem[Gratton et al.(2003)]{gra03}
Gratton, R. G. et al. 2003, \aap, 404, 187

\bibitem[Greggio(1996)]{gre96} 
Greggio, L. 1996, in The Interplay between Massive Star Formation, the ISM, and Galaxy Evolution, ed. D. Kunth et al. (Gif-suf-Yvette: Edition Fronti{\'e}res), 98

\bibitem[Greggio(2005)]{gre05} 
Greggio, L. 2005, \aap, 441, 1055

\bibitem[Greggio \& Renzini(1983)]{gre83} 
Greggio, L. \& Renzini, A. 1983, \aap, 118, 217

\bibitem[Hachisu \& Kato(2003ab)]{hac03} 
Hachisu, I., \& Kato, M. 2003a, \apj, 590, 445

\bibitem[Hachisu \& Kato(2003b)]{hac03b} 
Hachisu, I., \& Kato, M. 2003b, \apj, 598, 527

\bibitem[Hachisu, Kato \& Nomoto(1996)]{hac96} 
Hachisu, I., Kato, M., \& Nomoto, K. 1996, \apj, 470, L97

\bibitem[Hachisu, Kato \& Nomoto(1999a)]{hac99} 
Hachisu, I., Kato, M., \& Nomoto, K. 1999a, \apj, 522, 487

\bibitem[Hachisu et al.(1999b)]{hac99u} 
Hachisu, I., Kato, M., Nomoto, K., \& Umeda, H. 1999b, \apj, 519, 314

\bibitem[Hachisu, Kato \& Nomoto(2008)]{hac08} 
Hachisu, I., Kato, M., \& Nomoto, K. 2008, \apj, 679, 1390 (HKN08)

\bibitem[Hamuy et al.(2003)]{ham03}
Hamuy, M., et al. 2003, Nature, 424, 651

\bibitem[Hamuy et al.(2000)]{ham00}
Hamuy, M., Trager, S. C., Pinto, P. A., Phillips, M. M., Schommer, R. A., Ivanov, V. \& Suntzeff, N. B. 2000, \aj, 120, 1479

\bibitem[Hamuy et al.(2001)]{ham01}
Hamuy, M., Trager, S. C., Pinto, P. A., Phillips, M. M., Schommer, R. A., Ivanov, V. \& Suntzeff, N. B. 2001, \aj, 122, 3506

\bibitem[Hamuy \& Pinto(1999)]{ham99}
Hamuy, M. \& Pinto, P. A. 1999, \aj, 117, 1185

\bibitem[Han et al.(1995)]{han95}
Han, Z., Eggleton, P. P., Podsiadlowski, P. \& Tout, C. A. 1995, \mnras, 277, 1443\

\bibitem[Hardin et al.(2000)]{har00}
Hardin, D. et al. 2000, \aap, 362, 419

\bibitem[Hillebrandt \& Niemeyer(2000)]{hil00}
Hillebrandt, W. \& Niemeyer, J. C. 2000, \araa, 38, 191

\bibitem[Hogeveen(1992)]{hog92}
Hogeveen, S. J. 1992, Ap\&SS, 196, 299

\bibitem[H\"{o}flich \& Khokhlov(1996)]{hof96}
H\"{o}flich, P., \& Khokhlov, A. 1996, \apj, 457, 500

\bibitem[Honda et al.(2004)]{hon04}
Honda, S. et al. 2004, \apj, 607, 474

\bibitem[Howell et al.(2009)]{how09}
Howell, D. A. et al. 2009, \apj, 691, 661

\bibitem[Iben \& Tutukov(1984)]{ibe84}
Iben, I. Jr., \& Tutukov, A. V. 1984, \apjs, 54, 335

\bibitem[Israelian et al.(1998)]{isr98}
Israelian, Garcia-Lopez, Rebolo 1998, \apj, 207, 805

\bibitem[Iwamoto \& Saio(1999)]{iwan99}
Iwamoto, N. \& Saio, H. 1999, \apj, 521, 297

\bibitem[Karakas \& Lattanzio(2007)]{kar07}
Karakas, A., \& Lattanzio, J. 2007, \pasa, 24, 103

\bibitem[Kirby et al.(2008)]{kir08}
Kirby, E. N., Simon, J. D., Geha, M., Guhathakurta, P., Frebel, A. 2008, \apj, 685, 43

\bibitem[Kobayashi(2003)]{kob03} 
Kobayashi, C., in "Elemental Abundances in Old Stars and Damped Lyman-$\alpha$ Systems", 25th General Assembly of the IAU - 2003, Joint Discussion 15, p. 21

\bibitem[Kobayashi(2008)]{kob08}
Proceedings of Science, 10th Symposium on Nuclei in the Cosmos, p.49-58

\bibitem[Kobayashi \& Arimoto(1999)]{kob99} 
Kobayashi, C., \& Arimoto, N. 1999, \apj, 527, 573

\bibitem[Kobayashi et al.(2007)]{kob07} 
Kobayashi, C., Springel, V, \& White, S. D. M. 2007, \mnras, 376, 1465

\bibitem[Kobayashi, Tsujimoto \& Nomoto(2000)]{kob00} 
Kobayashi, C., Tsujimoto, T., \& Nomoto, K. 2000, \apj, 539, 26 (K00)

\bibitem[Kobayashi et al.(1998)]{kob98} 
Kobayashi, C., Tsujimoto, T., Nomoto, K., Hachisu, I, \& Kato, M. 1998, \apj, 503, L155 (K98)
 
\bibitem[Kobayashi et al.(2006)]{kob06} 
Kobayashi, C., Umeda, H., Nomoto, K., Tominaga, N., \& Ohkubo, T. 2006, \apj, 653, 1145 (K06)

\bibitem[Koch et al.(2008)]{koc08}
Koch, A. et al. 2008, \aj, 135, 1580

\bibitem[Kodama \& Arimoto(1997)]{kod97} 
Kodama, T., \& Arimoto, N., 1997, \aap, 320, 41

\bibitem[Kouwenhoven et al.(2005)]{kou05}
Kouwenhoven, M. B. N., Brown, A. G. A., Zinnecker, H., Kaper, L., \& Portegies Zwart, S. F. 2005, \aap, 430, 137

\bibitem[Kroupa(1995)]{kro95}
Kroupa, P. 1995, \mnras, 277, 1491

\bibitem[Kroupa(2008)]{kro07}
Kroupa, P. 2008, ASP Conference Series, 390, 3

\bibitem[Kuznetsova et al.(2008)]{kuz07}
Kuznetsova, N. et al. 2008, \apj, 673, 981

\bibitem[Livio(2001)]{liv01}
Livio, M. 2001, in Supernovae and Gamma-Ray Bursts, eds. M. Livio et al. (New York: Cambridge University Press), 334

\bibitem[Madgwick et al.(2003)]{mad03}
Madgwick, D. S., Hewett, P. C., Mortlock, D. J., \& Wang, L. 2003, \apj, 599, L33

\bibitem[Mannucci et al.(2005)]{man05}
Mannucci, F. et al. 2005, \aap, 433, 807

\bibitem[Mannucci et al.(2006)]{man06}
Mannucci, F., Della Valle, M., \& Panagia, N. 2006, \mnras, 370, 773

\bibitem[Maoz(2008)]{mao08}
Maoz, D., 2008, \mnras, 384, 267

\bibitem[Matteucci(1997)]{mat97}
Matteucci, F. 1997, in ASP Conference Series, Vol. 126, From Quantum Fluctuations to Cosmological Structures, 495

\bibitem[Matteucci(2001)]{matteucci2001}
Matteucci, F. 2001, The Chemical Evolution of the Galaxy (Kluwer Academic Pub.)

\bibitem[Matteucci \& Greggio(1986)]{mat86}
Matteucci, F., \& Greggio, L. 1986, \aap, 154, 279

\bibitem[Matteucci et al.(2006)]{mat06}
Matteucci, F., Panagia, N., Pipino, A., Mannucci, F., Recchi, S., \& Della Valle, M. 2006, \mnras, 372, 265

\bibitem[Matteucci \& Recchi(2001)]{mat01} 
Matteucci, F. \& Recchi, S. 2001, \apj, 558, 351 (MR01)

\bibitem[McWilliam et al.(1995)]{mcw95}
McWilliam, A., Preston, G.W., Sneden, C., \& Searle, L. 1995,
AJ, 109, 2757

\bibitem[Napiwotzki(2007)]{nap07}
Napiwotzki, R. 2007 talk at the KITP workshop on Paths to Exploding Stars: Accretion and Eruption, Santa Barbara, 19-23 March, http://online.kitp.ucsb.edu/online/snovae\_c07

\bibitem[Neil et al.(2006)]{nei06}
Neil, J. D. et al. 2006, \apj, 132, 1126

\bibitem[Nissen et al.(2004)]{nis04}
Nissen, P. E., Chen, Y. Q., Asplund, M., \& Pettini, M 2004, \aap, 415, 993

\bibitem[Nissen et al.(2007)]{nis07}
Nissen, P. E., Akerman, C., Asplund, M., Fabbian, D., Kerber, F., K\"aufl, H. U., \& Pettini, M. 2007, \aap, 469, 319

\bibitem[Nomoto, Iwamoto \& Kishimoto(1997)]{nom97a}
Nomoto, K., Iwamoto, K., \& Kishimoto, N. 1997, Science, 276, 1378

\bibitem[Nomoto et al.(2007)]{nom07}
Nomoto, K., Saio, H., Hachisu, I. \& Kato, M. 2007, 663, 1269

\bibitem[Nomoto et al.(2006)]{nom06}
Nomoto, K., Tominaga, N., Umeda, H., Kobayashi, C, \& Maeda, K. 2006, Nuclear Physics A, 777, 424

\bibitem[Nomoto et al.(2000)]{nom00a}
Nomoto, K., Umeda, H., Hachisu, I., Kako, M., Kobayashi, C., \& Tsujimoto, T. 2000, in Type Ia Supernovae: Theory and Cosmology, eds. J. Niemeyer \& J. Truran (New York: Cambridge University Press), 63 (asSpolaortro-ph/9907386)

\bibitem[Nomoto et al.(1994)]{nom94}
Nomoto, K., Yamaoka, H., Shigeyama, T., Kumagai, S., \& Tsujimoto, T.
1994, in Supernovae, Les Houches Session LIV, ed. S. A. Bludman et al.
(Amsterdam: North-Holland), 199

\bibitem[Nordstr\"{o}m et al.(2004)]{nord04}
Nordstr\"{o}m, B., Mayor, M., Andersen, J., Holmberg, J., Pont, F., J\/{o}rgensen, B. R., Olsen, E. H., Udry, S., Mowlavi, N. 2004, \aap, 418, 989

\bibitem[Nugent et al.(1997)]{nug97}
Nugent, P., Baron, E., Branch, D., Fisher, A., \& Hauschildt, P. H. 1997, 
\apj, 485, 812

\bibitem[Pagel(1997)]{pagel1997} 
Pagel, B. E. J. 1997, Nucleosynthesis and Chemical Evolution of Galaxies (Cambridge Unv. Press)

\bibitem[Pain et al.(1996)]{pai96} 
Pain, R., et al. 1996, \apj, 473, 356

\bibitem[Pain et al.(2002)]{pai02}
Pain, R., et al. 2002, \apj, 577, 120

\bibitem[Patat et al.(2007)]{pat07}
Patat, F. et al. 2007, Science, 317, 924 

\bibitem[Pettini et al.(1999)]{pet99}
Pettini, M. Ellison, S. L., Steidel, C. C., \& Bowen, D. V. 1999, \apj, 510, 576

\bibitem[Portegies Zwart(1995)]{por95}
Portegies Zwart, S. F. 1995, \aap, 296, 691

\bibitem[Poznanski et al.(2007)]{poz07}
Poznanski, D. et al. 2007, \mnras, 382, 1169

\bibitem[Prieto, Stanek, \& Beacom(2008)]{pri08}
Prieto, J. L., Stanek, K. Z., \& Beacom, J. F. 2008, \apj, 673, 999

\bibitem[Primas et al.(2000)]{pri00}
Primas, F., Brugamyer, E., Sneden, C., King, J. R., Beers, T. C., Boesgaard, A. M., Deliyannis, C. P. 2000, in The First Stars, eds. A. Weiss, T. Abel, \& V. Hill (Berlin: Springer), 51

\bibitem[Prochaska \& Wolfe(2002)]{pro02}
Prochaska \& Wolfe(2002), \apj, 566, 68

\bibitem[Quimby et al.(2006)]{qui06}
Quimby, R. et al. 2006, \apj, 636, 400

\bibitem[Riess(1996)]{rie96}
Riess, A. 1996, Ph.D thesis, Harvard University

\bibitem[Roberts \& Haynes(1994)]{rob94}
Roberts, M. S. \& Haynes, M. P. 1994, \araa, 32, 115

\bibitem[Ruiz-Lapuente, Burkert, \& Canal(1995)]{rui95}
Ruiz-Lapuente, P., Burkert, A. \& Canal, R. 1995, \apj, 447, L69

\bibitem[Ryan et al.(1996)]{rya96}
Ryan, S. G., Norris, J. E., \& Beers, T. C. 1996, \apj, 471, 254

\bibitem[Saio \& Nomoto(1985)]{sai85}
Saio, H., \& Nomoto, K. 1985, \aap, 150, L21

\bibitem[Saio \& Nomoto(1998)]{sai98}
Saio, H., \& Nomoto, K. 1998, \apj, 500, 388

\bibitem[Scannapieco \& Bildsten(2005)]{sca05}
Scannapieco, E. \& Bildsten, L. 2005, \apj, 629, L85

\bibitem[\protect\citeauthoryear{Schiminovich et al.}{2005}]{sch05}
Schiminovich, D. et al. 2005, \apj, 619, L47

\bibitem[Shatsky \& Tokovinin(2002)]{sha02}
Shatsky, N. \& Tokovinin, A., 2002, \aap, 382, 92

\bibitem[Shetrone, C{\^o}t{\'e} \& Sargent(2001)]{she01} 
Shetrone, M. C{\^o}t{\'e}, P., \& Sargent, W. L. W. 2001, \apj, 548, 592

\bibitem[Shetrone et al.(2003)]{she03} 
Shetrone, M. Venn, K. A., Tolstoy, E., Primas, F., Hill, V., \& Kaufer, A. 2003, \aj, 125, 684

\bibitem[Simon et al.(2007)]{sim07}
Simon, J. D. et al. 2007, \apj, 671, L25

\bibitem[Sneden et al.(1991)]{sne91}
Sneden, C., Gratton, R. G., \& Crocker, D. A. 1991, \aap, 246, 354

\bibitem[Spolaor et al.(2009)]{spr09}
Spolaor, M., Kobayashi, C., Forbes, D. A., Couch, W. J., \& Hau, G. K., 2009, \mnras, submitted

\bibitem[Strolger et al.(2002)]{str02}
Strolger, L.-G. et al. 2002, \apj, 124, 2905

\bibitem[Strolger et al.(2004)]{str04}
Strolger, L.-G. et al. 2004, \apj, 613, 200

\bibitem[Sullivan et al.(2006)]{sul06}
Sullivan, M. et al. 2006, \apj, 648, 868

\bibitem[Thomas \& Maraston(2003)]{tho03}
Thomas, D. \& Maraston, C. 2003, \aap, 401, 429

\bibitem[Tohline(2002)]{toh02}
Tohline, J. E. 2002, \araa, 40, 349

\bibitem[Tolstoy et al.(2003)]{tol03}
Tolstoy et al. 2003, \aj, 125, 707

\bibitem[Tonry et al.(2003)]{ton03}
Tonry, J. L. et al. 2003, \apj, 594, 1

\bibitem[Tovmassian et al.(2008)]{tov07}
Tovmassian, G., Tomsick, J., Napiwotzki, R., Yungelson, L., 
Stasi\'nska, G., Pen\~a, M., \& Richer, M. 2008, ASP Conference Series, 968, 62

\bibitem[Tremonti et al.(2004)]{tre04}
Tremonti, C. A. et al. 2004, \apj, 613, 898

\bibitem[Tutukov \& Yungelson(1994)]{tut94} 
Tutukov, A. V., \& Yungelson, L. R. 1994, \mnras, 268, 871

\bibitem[Tutukov \& Fedorova(2007)]{tut07} 
Tutukov, A. V., \& Fedorova, A. V. 2007, Astronomy Reports, Vol. 51, p.291

\bibitem[Umeda \& Nomoto(2002)]{ume02}
Umeda, H. \& Nomoto, K. 2002, \apj, 565, 385

\bibitem[Venn et al.(2004)]{ven04}
Venn, K. A., Irwin, M., Shetrone, M. D., Tout, C. A., Hill, V., Tolstoy, E. 2004, \aj, 128, 1177

\bibitem[Webbink(1984)]{web84}
Webbink, R. F. 1984, \apj, 277, 355

\bibitem[Wyse \& Gilmore(1995)]{wys95}
Wyse, R. F. G., \& Gilmore, G. 1995, \aj, 110, 2771

\bibitem[Yoshii, Tsujimoto, \& Kawara(1998)]{yos98}
Yoshii, Y., Tsujimoto, T., \& Kawara, K. 1998, \apj, 507, 113

\bibitem[Yungelson(2005)]{yun05}
Yungelson, L. R. 2005, in: White dwarfs: cosmological and galactic probes, eds. E. M. Sion, S. Vennes \& H. L. Shipman (Dordrecht: Springer), 163

\bibitem[Yungelson \& Livio(1998)]{yun98}
Yungelson, L., \& Livio, M. 1998, \apj, 497, 168

\end{thebibliography}
\end{document}